\documentstyle[aps,epsfig,floats]{revtex}

\tighten
\renewcommand{\ol}{\overline}
\newcommand{\Pslash}{\kern 0.2 em P\kern -0.56em \raisebox{0.3ex}{/}}
\newcommand{\pslash}{\kern 0.2 em p\kern -0.4em /}
\newcommand{\kslash}{\kern 0.2 em k\kern -0.45em /}
\newcommand{\Sslash}{\kern 0.2 em S\kern -0.56em \raisebox{0.3ex}{/}}
\newcommand{\Mslash}{\kern 0.2 em M\kern -0.70em \raisebox{0.3ex}{/}}
\newcommand{\g}{\gamma}
\newcommand{\sig}{\sigma}
\newcommand{\eps}{\epsilon}

\newcommand{\one}{1\hspace{-2pt}\rule[0.08ex]{0.45pt}{1.5ex}\hspace{2pt}}

\newcommand{\lcvec}[3]{\left[\;#1\;,\;#2\;,\;#3\;\right]}
\newcommand{\sT}{{\scriptscriptstyle T}}
\newcommand{\nn}{\nonumber}
\newcommand{\newangle}{{<\kern -0.3 em{\scriptscriptstyle )}}}
\newcommand{\sumint}{\kern 0.2 em {\textstyle\sum} \kern -1.1 em \int_X}

\begin{document}
\draft
\title{
\begin{flushright}
{\small hep-ph/9907488}
\end{flushright}
Two-hadron interference fragmentation functions \\
Part II: a model calculation
}

\author{A.~Bianconi$^{1}$, S.~Boffi$^{2}$, R. Jakob$^{2}$, M.~Radici$^{2}$}
\address{
1. Dipartimento di Chimica e Fisica per i Materiali e per l'Ingegneria,\\ 
Universit\`{a} di Brescia, I-25133 Brescia, Italy,\\
2. Dipartimento di Fisica Nucleare e Teorica, Universit\`{a} di Pavia, 
and\\
Istituto Nazionale di Fisica Nucleare, Sezione di Pavia, 
I-27100 Pavia, Italy}

\date{July 26, 1999}

\maketitle

\begin{abstract}
We present a model calculation of leading order interference fragmentation 
functions arising from the distribution of two hadrons produced in the same 
jet of a fragmenting quark in a hard process. Using a simple spectator model
for the mechanism of the hadronization process, for the first time we give a
complete calculation and numerical 
estimates for the example of a proton-pion pair production with its 
invariant mass on the Roper resonance.  
\end{abstract}


\section{Introduction}\label{sec:intro}
In a companion paper~\cite{noi} we have discussed the general framework of 
two-hadron interference fragmentation functions, their definitions and
properties. We have given the cross section of two-hadron inclusive 
lepton-nucleon scattering as an example how they can be disentangled 
experimentally by using azimuthal 
angular dependences of a hard process. In the present article we supplement
our investigation with a spectator model calculation of the 
interference fragmentation functions for the special case of the 
hadron-pair in the quark jet being a pion and 
a proton. 

Fragmentation functions (FF), like distribution 
functions (DF), ultimately contain the complete
information on the confinement of partons inside hadrons. Factorization
theorems for hard processes, whenever available, ensure their universality, 
i.e.\ FF (and DF) are process independent; once determined in a hard process 
they can be used to predict observables in other hard reactions.
FF and DF describe the properties of partons in hadronic bound states
at a low scale. Their intrinsic non-perturbative nature presently 
prevents a complete  
calculation from first principles within the framework of Quantum 
Chromodynamics. On the experimental side, several of the DF are nowadays 
quite well 
known from deep inelastic inclusive lepton-nucleon scattering, the Drell-Yan 
process or from  jet production in nucleon-nucleon collisions. The 
experimental knowledge of the various FF, however, is very 
poor because it requires the ability of
measuring more exclusive channels in hard processes (semi-inclusive
cross sections, transverse momenta and/or polarizations of detected
hadrons, etc..). Therefore, model calculations of FF are of particular
importance to learn about their possible features and to estimate cross 
sections for future experiments. 

When new channels are open in the final state, several FF arise which 
can also be time-reversal odd~\cite{vari1,danielpiet} (for brevity, 
``T-odd''), in the sense that the
constraints due to time-reversal invariance cannot be applied. The main
reason for that is the existence of residual Final State Interactions
(FSI) between the produced hadron(s) and the remnants of the fragmenting
quark.  Thus, FF not only contain 
information complementary to the one given by DF, but in particular
the T-odd ones also represent a 
powerful tool to explore processes taking place inside the 
jet. Moreover, specific T-odd interference FF can represent the necessary
chiral odd partner to address the transversity 
distribution $h_1$~\cite{coll}. Therefore, the aim of the present paper is 
to calculate T-odd interference FF within a model.

The presence of FSI means that in the fragmentation process there 
are at least two leading interfering channels. In the previous 
paper~\cite{noi} we have shown that the interference of
two channels is not enough to generate
``T-odd'' FF. For example, in the case of one-hadron semi-inclusive
processes we can model FSI by introducing an external potential. Despite 
this simplistic approach, we encounter a serious mathematical 
difficulty, because the
potential in principle breaks the translational and rotational invariance
of the problem. Simplifying assumptions about the symmetry of the
potential can be introduced, but at the price of loosing any contribution
to the ``T-odd'' structure of the amplitude. Alternatively, we have
explicitly shown that a genuine difference in the Lorentz structure of
the vertices describing the two interfering processes is needed to produce
a ``T-odd'' contribution~\cite{noi}. All this means that addressing 
``T-odd'' FF in 
one-hadron semi-inclusive processes requires the ability of describing the
FSI between the outgoing hadron and the remnant of the jet, relating the 
modifications of the hadron wave function to a realistic microscopic 
description of the fragments. 

Because of these arguments, a more convenient way to investigate T-odd FF 
is to look at the
hadronization of a current quark where two leading hadrons are detected
within the same jet, considering the remnant of the jet as a spectator 
and summing over all its possible configurations. By interference of 
different channels producing the two hadrons, FF emerge which are 
``T-odd'', and can be both chiral even or chiral odd~\cite{noi}. 
For the case of the two hadrons 
being a pair of pions the resulting FF have been proposed to investigate 
the transverse spin dependence of fragmentation. Collins and 
Ladinsky~\cite{collad} considered the interference of a scalar resonance 
with the channel of independent successive two pion production. Jaffe, Jin 
and Tang~\cite{jafjitang} proposed the interference of $s$- and $p$-wave 
production channels, where the relevant phase shifts are essentially 
known. We will estimate the FF in the case of the pair being a proton and 
a pion produced either through non-resonant channels or 
through the Roper ($1440$ MeV) resonance. 

This paper is organized as 
follows. In Sec.~\ref{sec:two} a brief
summary of the definitions, properties and relevant formulae about 
two-hadron FF are given together with a description of the 
kinematics. We restrict ourselves to the information necessary to keep this
paper self-contained; full details can be found 
in Ref.~\cite{noi}. In Sec.~\ref{sec:three} we
discuss an extended version of the spectator model used in
Ref.~\cite{spectator,melni,jak97} and here adopted for the calculation. In 
Sec.~\ref{sec:four} numerical estimates are presented and discussed for
``T-odd'' FF that emerge to leading order. Finally, a brief summary is
given in Sec.~\ref{sec:summary}.

\section{Fragmentation functions for proton-pion pair production}
\label{sec:two}

\begin{figure}[h]
\begin{center}
\psfig{file=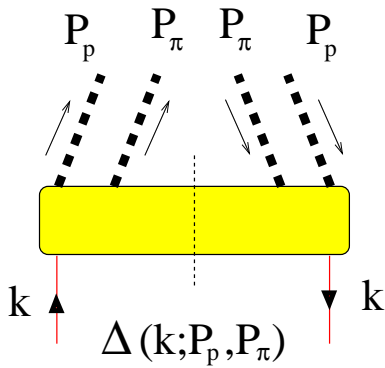, width=5cm}
\end{center}
\caption{Quark-quark correlation function for the fragmentation of a quark 
into a proton-pion pair.}
\label{Delta-2h}
\end{figure}

As discussed in some detail in the companion paper~\cite{noi} the two-hadron
FF can be defined as Dirac projections of (partly integrated) quark-quark
correlation functions. In the field theoretical description of hard 
processes those functions  describe the soft parts which connect quark 
lines to hadrons, i.e.\  they are hadronic matrix elements of non-local 
operators built from quark (or gluon) fields. For a quark 
fragmenting into a jet which contains a pion-proton pair the
appropriate correlation function (in a light-cone gauge) is 
\begin{equation}
\Delta_{ij}(k;P_p,P_\pi)= \sumint \; \int
\frac{d^{\,4\!}\zeta}{(2\pi)^4} \; e^{ik\cdot\zeta}\;
\langle 0|\psi_i(\zeta)|\pi,p,X\rangle
\langle X,p,\pi|\ol{\psi}_j(0)|0\rangle
\label{eq:defDelta}
\end{equation}
where the sum runs over all the possible intermediate states containing the
pair. The momenta are indicated in Fig.~\ref{Delta-2h}.

Before proceeding with the definition of FF obtained from $\Delta$, we 
summarize the discussion of kinematics as given in~\cite{noi}. For 
clarity we give an explicit parametrization in a 
reference frame where
the sum of proton and pion momenta $P_h=P_p+P_\pi$ has no transverse
components. Two dimensionless light-like vectors $n_+$ and $n_-$ satisfying 
$n_+\cdot n_-=1$ can be used to decompose a 
generic 4-vector $a$ in its light-cone components 
$a^\pm=\left(a^0\pm a^3\right)/\sqrt{2}=a_\cdot n_\mp$ and a 
two-dimensional transverse vector $\vec a_\sT$. Throughout the paper we use 
the notation $a^\mu=\left[a^-,a^+, {\vec a}_\sT\right]$. A possible
parametrization of the momenta is
\begin{eqnarray}
k&=&
\lcvec{\frac{P_h^-}{z_h}}{z_h\frac{k^2+{\vec k}_T^2}{2P_h^-}}{{\vec k}_T} 
\;,\nn\\
P_p&=&
\lcvec{\xi\,P_h^-}{\frac{M_p^2+{\vec R}_T^2}{2\,\xi\,P_h^-}}{{\vec R}_T}
\;,\nn\\
P_\pi&=&
\lcvec{(1-\xi)\,P_h^-}{\frac{M_\pi^2+{\vec R}_T^2}{2\,(1-\xi)\,P_h^-}}
{-{\vec R}_T} \;,
\label{eq:vectors2}
\end{eqnarray}
where $R\equiv(P_p-P_\pi)/2$ is (half of) the relative momentum 
between the proton and the pion, and 
we defined the light-cone momentum fractions
\begin{equation}
z_h=\frac{{P_h}^-}{k^-}\;,\qquad
\xi=\frac{P_p^-}{P_h^-}\;,\qquad
(1-\xi)=\frac{P_\pi^-}{P_h^-} \;.
\label{eq:z}
\end{equation} 
With the constraint of reproducing on-shell hadrons at fixed mass 
$(P_p^2=M_p^2, P_\pi^2=M_\pi^2)$, the dependence of $\Delta$ on the momenta 
$k,P_p,P_\pi$ in Eq.~(\ref{eq:defDelta}) can be reexpressed in terms of
seven independent variables: the light-cone 
component of the hadron pair momentum, $P_h^-$; the light-cone fraction 
of the quark momentum carried by the hadron pair, $z_h=P_h^-/k^-$; 
the fraction of hadron pair momentum carried by each individual hadron, 
$\xi$; the four independent invariants that can 
be formed in this case, i.e.
\begin{mathletters}
\label{explinv2}
\begin{eqnarray}
\tau_h &=& k^2 \;,\nn \\
\sig_h &=& 2k\cdot P_h =\left\{\frac{M_p^2+{\vec R}_T^2}{z_h\,\xi}
+\frac{M_\pi^2+{\vec R}_T^2}{z_h\,(1-\xi)}\right\}+z_h\,(\tau_h+{\vec k}_T^2)
\;,\label{eq:sigh2} \\
\sig_d &=& 2k\cdot (P_p-P_\pi) =\left\{\frac{M_p^2+{\vec R}_T^2}{z_h\,\xi}
-\frac{M_\pi^2+{\vec R}_T^2}{z_h\,(1-\xi)}\right\}+z_h(2\xi-1)(\tau_h+
{\vec k}_T^2)-4\,{\vec k}_T\cdot{\vec R}_T \;, \label{eq:sigd2} \\
M_h^2 &=& P_h^2 = 2\,P_h^+\,P_h^-=\left\{\frac{M_p^2+{\vec R}_T^2}{\xi}
+\frac{M_\pi^2+{\vec R}_T^2}{1-\xi}\right\} \;, 
\label{eq:mh2}
\end{eqnarray}
\end{mathletters}
where Eq.\ (\ref{eq:mh2}) can also be expressed as
\begin{equation}
{\vec R}_T^2=\xi\,(1-\xi)\,M_h^2-(1-\xi)\,M_p^2-\xi\,M_\pi^2 \; .
\label{eq:pt2}
\end{equation}
In the following definitions of the FF the correlation 
function $\Delta$ will occur integrated over the (hard-scale) 
suppressed light-cone component $k^+$ at $k^-=P_h^-/z_h$ and traced with a 
certain Dirac structure $\Gamma$ in the form
\begin{equation} 
\Delta^{[\Gamma]} = \frac{1}{4z_h} \int dk^+\; \int dk^-\; 
\delta\left(k^--\frac{P_h^-}{z_h}\right)\; \mbox{Tr}[\Delta\Gamma] \; .
\label{eq:projDelta}
\end{equation}
This will reduce the number of independent variables 
to the five $z_h,\xi,{\vec k}_\sT^{\,2},M_h^2,\sig_d$. 
To make this evident we rewrite the integration in a covariant
way using
\begin{equation}
2\,P_h^-=\frac{d\sig_h}{dk^+},
\qquad
2\,k^+=\frac{d\tau_h}{dk^-} ,
\end{equation} 
and the relation
\begin{equation} 
\frac{1}{2k^+}\delta\left(k^--\frac{P_h^-}{z_h}\right)=
\delta\left(2k^+k^--\frac{2k^+P_h^-}{z_h}\right)=
\delta\left(\tau_h+{\vec k}_\sT^{\,2}-\frac{\sig_h}{z_h}
+\frac{M_h^2}{z_h^2}\right)  \; .
\end{equation} 
We get the final expression
\begin{equation} 
\Delta^{[\Gamma]}(z_h,\xi,{\vec k}_\sT^{\,2},M_h^2,\sig_d)=
\int d\sig_h \, d\tau_h \;\delta\left(\tau_h+{\vec k}_\sT^{\,2}-
\frac{\sig_h}{z_h}+\frac{M_h^2}{z_h^2}\right)\;\frac{\mbox{Tr}[
\Delta(z_h,\xi,P_h^-,\tau_h,\sig_h,M_h^2,\sig_d) \; \Gamma]}{8z_hP_h^-} 
\;,
\label{eq:projDelta2}
\end{equation}
where the dependence on the transverse quark momentum ${\vec k}_\sT^{\,2}$ 
through $\sig_h$ is made explicit by means of Eqs. (\ref{eq:sigh2}) and 
(\ref{eq:pt2}). Using the parametrization Eq.\ (\ref{explinv2}) it is 
possible to 
further reexpress the set of kinematical variables as $z_h$, $\xi$, 
${\vec k}_\sT^{\,2}$ and 
${\vec R}_\sT^{\,2}$,  ${\vec k}_\sT \cdot {\vec R}_\sT$, 
where ${\vec R}_\sT$ is (half of) the transverse momentum between the two 
hadrons in the considered frame. 

\begin{figure}[h]
\begin{center}
\psfig{file=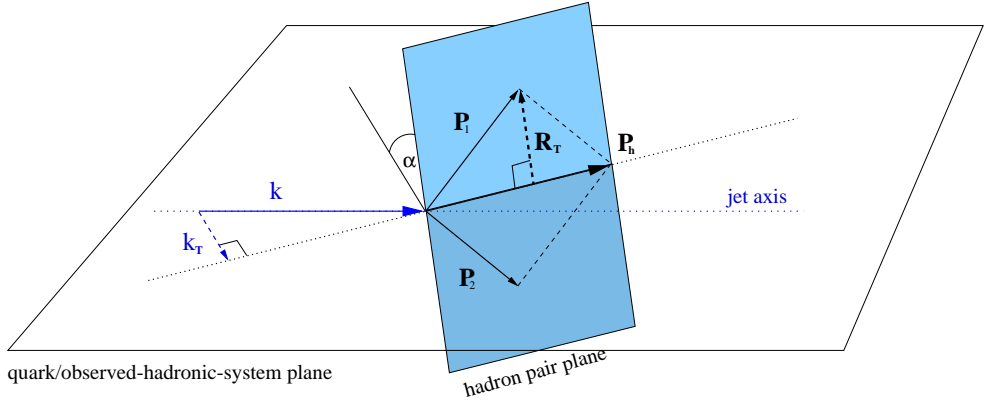, width=13cm}
\end{center}
\caption{Kinematics for a fragmenting quark jet containing a pair of 
leading hadrons.}
\label{fig2}
\end{figure}

Two-hadron FF to leading order are defined by the following Dirac projections 
\begin{eqnarray} 
\Delta^{[\g^-]}(z_h,\xi,{\vec k}_\sT^{\,2},{\vec R}_\sT^{\,2},
   {\vec k}_\sT \cdot {\vec R}_\sT) &=& 
D_1(z_h,\xi,{\vec k}_\sT^{\,2},{\vec R}_\sT^{\,2},
   {\vec k}_\sT \cdot {\vec R}_\sT)   
\label{eq:projg-} \\[2mm]
\Delta^{[\g^- \g_5]}(z_h,\xi,{\vec k}_\sT^{\,2},{\vec R}_\sT^{\,2},
   {\vec k}_\sT \cdot {\vec R}_\sT) &=& 
\frac{\eps_\sT^{ij} \,R_{\sT i}\,k_{\sT j}}{M_1\,M_2}\;
G_1^\perp (z_h,\xi,{\vec k}_\sT^{\,2},{\vec R}_\sT^{\,2},
   {\vec k}_\sT \cdot {\vec R}_\sT) 
\label{eq:projg-g5} \\[2mm]
\Delta^{[i\sig^{i-} \g_5]}(z_h,\xi,{\vec k}_\sT^{\,2},{\vec R}_\sT^{\,2},
   {\vec k}_\sT \cdot {\vec R}_\sT) &=& 
{\epsilon_\sT^{ij}R_{\sT j}\over M_1+M_2}\, 
H_1^{\newangle}(z_h,\xi,{\vec k}_\sT^{\,2},{\vec R}_\sT^{\,2},
   {\vec k}_\sT \cdot {\vec R}_\sT)\nn\\
&&{}+{\epsilon_\sT^{ij}k_{\sT j}\over M_1+M_2}\,
H_1^\perp(z_h,\xi,{\vec k}_\sT^{\,2},{\vec R}_\sT^{\,2},
   {\vec k}_\sT \cdot {\vec R}_\sT) 
 \; , \label{eq:projsigi-g5}
\end{eqnarray}
where we used transverse 4-vectors defined as 
$a_\sT^\mu=g_\sT^{\mu\nu}\,a_\nu=[0,0,\vec a_\sT]$ (with 
$g_\sT^{\mu\nu}=g^{\mu\nu}-n_+^\mu n_-^\nu-n_+^\nu n_-^\mu$).
In this manner, the FF 
depend on how much of the fragmenting quark momentum is carried by the 
hadron pair $(z_h)$, on the way this momentum is shared inside the pair 
$(\xi)$, and on the ``geometry'' of the pair, namely on the relative 
momentum of the two hadrons $({\vec R}_\sT^{\,2})$ and on the relative 
orientation between the pair plane and the quark jet axis 
(${\vec k}_\sT^{\,2}$, ${\vec k}_\sT \cdot {\vec R}_\sT$, see also 
Fig.~\ref{fig2}).

In Eqs.~(\ref{eq:projg-})-(\ref{eq:projsigi-g5}), 
$D_1, G_1^{\perp}, H_1^{\perp}, H_1^{\newangle}$ are the interference FF that
arise to leading order in 
$1/Q$ for the fragmentation of a current quark into two 
unpolarized hadrons inside the same jet, in the case under consideration a
proton-pion pair. The different Dirac structures in the projections are 
related to different spin states of 
the fragmenting quark and lead to a nice probabilistic
interpretation~\cite{noi}:
$D_1$ is the probability for an unpolarized quark to 
produce a pair of unpolarized hadrons; $G_1^\perp$ is the difference of 
probabilities for a longitudinally polarized quark with opposite 
chiralities to produce a pair of unpolarized hadrons; $H_1^{\newangle}$ 
and $H_1^\perp$ both are differences of probabilities for a transversely 
polarized quark with opposite spins to produce a pair of unpolarized 
hadrons~\cite{noi}. $D_1$ is chiral even and ``T-even''; instead,
$G_1^{\perp}, H_1^{\perp}, H_1^{\newangle}$ are (naive) ``T-odd''.
$G_1^{\perp}$ is also chiral even, while $H_1^{\perp}, H_1^{\newangle}$
are chiral odd and, as pointed out in Ref.~\cite{noi}, represent the natural 
partner in a measurement of the transversity distribution $h_1$, in a 
a sort of ``double'' Collins effect~\cite{coll}.

\section{Spectator model}
\label{sec:three}

In this section we extend the formalism of the so-called diquark
spectator model~\cite{spectator,melni,jak97} to calculation of 
the two-hadron interference FF, specializing it to the emission of a
proton-pion pair. 

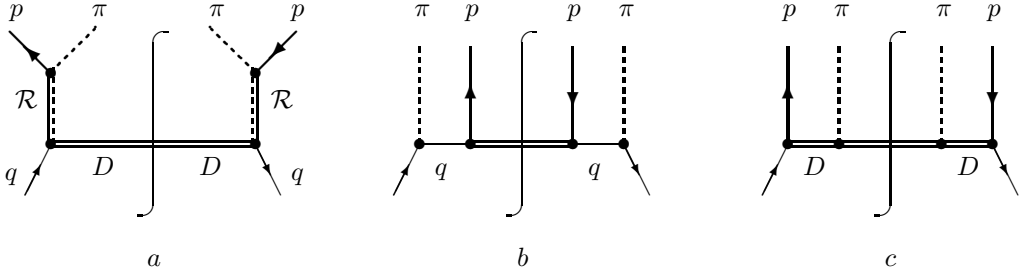
\begin{figure}[hbtp]
\begin{center}
\unitlength=0.68mm
%
%
\begin{picture}(60,45)
\put(10,15){\circle*{2}} 
\put(5,5){\line(1,2){5}}\put(5,5){\vector(1,2){3}} 
\multiput(10,14.5)(0,1){2}{\line(1,0){20}} 
\put(10,29){\circle*{2}} 
\put(2,40){$p$}\put(18,40){$\pi$} 
\put(1,8){$q$}\put(3,22){${\cal R}$}\put(18,9){$D$} 
\thicklines
 \put(10,29){\line(-1,1){8}}\put(10,29){\vector(-1,1){5}} 
 \multiput(10.0,29.0)(1.4,1.4){7}{\circle*{0.6}} 
 \multiput(10.2,29.2)(1.4,1.4){7}{\circle*{0.6}} 
 \multiput(10.4,29.4)(1.4,1.4){7}{\circle*{0.6}} 
 \put(9.5,15){\line(0,1){14}}              
 \multiput(10.5,15)(0,2){8}{\line(0,1){1}} 
\thinlines
\put(30,4){\line(0,1){30}}
\put(27,4){\oval(6,6)[br]}
\put(33,34){\oval(6,6)[tl]}
\put(50,15){\circle*{2}} 
\put(50,15){\line(1,-2){5}}\put(50,15){\vector(1,-2){3}} 
\multiput(30,14.5)(0,1){2}{\line(1,0){20}} 
\put(50,29){\circle*{2}} 
\put(57,40){$p$}\put(41,40){$\pi$} 
\put(57,8){$q$}\put(53,22){${\cal R}$}\put(39,9){$D$} 
\thicklines
 \put(50,29){\line(1,1){8}}\put(58,37){\vector(-1,-1){5}} 
 \multiput(50.0,29.0)(-1.4,1.4){7}{\circle*{0.6}} 
 \multiput(49.8,29.2)(-1.4,1.4){7}{\circle*{0.6}} 
 \multiput(49.6,29.4)(-1.4,1.4){7}{\circle*{0.6}} 
 \put(50.5,15){\line(0,1){14}}              
 \multiput(49.5,15)(0,2){8}{\line(0,1){1}} 
\thinlines
\end{picture} \qquad
%
%
\begin{picture}(60,45)
\put(10,15){\circle*{2}} 
\put(5,5){\line(1,2){5}}\put(5,5){\vector(1,2){3}} 
\put(20,15){\circle*{2}} 
\multiput(20,14.5)(0,1){2}{\line(1,0){10}} 
\put(10,15){\line(1,0){10}} 
\put(19,40){$p$}\put(9,40){$\pi$}\put(13,9){$q$} 
\thicklines
 \multiput(10,15)(0,2){10}{\line(0,1){1}} 
 \put(20,15){\line(0,1){19}}\put(20,15){\vector(0,1){12}} 
\thinlines
\put(30,4){\line(0,1){30}}
\put(27,4){\oval(6,6)[br]}
\put(33,34){\oval(6,6)[tl]}
\put(50,15){\circle*{2}} 
\put(55,5){\line(-1,2){5}}\put(50,15){\vector(1,-2){3}} 
\put(40,15){\circle*{2}} 
\multiput(30,14.5)(0,1){2}{\line(1,0){10}} 
\put(40,15){\line(1,0){10}} 
\put(39,40){$p$}\put(49,40){$\pi$}\put(43,9){$q$} 
\thicklines
 \multiput(50,15)(0,2){10}{\line(0,1){1}} 
 \put(40,15){\line(0,1){19}}\put(40,34){\vector(0,-1){12}} 
\thinlines
\end{picture} \qquad
%
%
\begin{picture}(60,45)
\put(10,15){\circle*{2}} 
\put(5,5){\line(1,2){5}}\put(5,5){\vector(1,2){3}} 
\put(20,15){\circle*{2}} 
\multiput(10,14.5)(0,1){2}{\line(1,0){20}} 
\put(19,40){$\pi$}\put(9,40){$p$}\put(13,9){$D$} 
\thicklines
 \multiput(20,15)(0,2){10}{\line(0,1){1}} 
 \put(10,15){\line(0,1){19}}\put(10,15){\vector(0,1){12}} 
\thinlines
\put(30,4){\line(0,1){30}}
\put(27,4){\oval(6,6)[br]}
\put(33,34){\oval(6,6)[tl]}
\put(50,15){\circle*{2}} 
\put(55,5){\line(-1,2){5}}\put(50,15){\vector(1,-2){3}} 
\put(40,15){\circle*{2}} 
\multiput(30,14.5)(0,1){2}{\line(1,0){20}} 
\put(39,40){$\pi$}\put(49,40){$p$}\put(43,9){$D$} 
\thicklines
 \multiput(40,15)(0,2){10}{\line(0,1){1}} 
 \put(50,15){\line(0,1){19}}\put(50,34){\vector(0,-1){12}} 
\thinlines
\end{picture}\\[2mm]
$a$ \hspace{45mm} $b$ \hspace{45mm} $c$
\end{center}
\caption{\label{fig3}Diagonal diagrams for quark $q$ decay into a proton
$p$ and a pion $\pi$ through a direct channel or a Roper resonance $R$.}
\end{figure}

\begin{figure}[hbtp]
\begin{center}
\unitlength=0.68mm
%
%
\begin{picture}(60,45)
\put(10,15){\circle*{2}} 
\put(5,5){\line(1,2){5}}\put(5,5){\vector(1,2){3}} 
\multiput(10,14.5)(0,1){2}{\line(1,0){20}} 
\put(10,29){\circle*{2}} 
\put(2,40){$p$}\put(18,40){$\pi$} 
\put(1,8){$q$}\put(3,22){${\cal R}$}\put(18,9){$D$} 
\thicklines
 \put(10,29){\line(-1,1){8}}\put(10,29){\vector(-1,1){5}} 
 \multiput(10.0,29.0)(1.4,1.4){7}{\circle*{0.6}} 
 \multiput(10.2,29.2)(1.4,1.4){7}{\circle*{0.6}} 
 \multiput(10.4,29.4)(1.4,1.4){7}{\circle*{0.6}} 
 \put(9.5,15){\line(0,1){14}}              
 \multiput(10.5,15)(0,2){8}{\line(0,1){1}} 
\thinlines
\put(30,4){\line(0,1){30}}
\put(27,4){\oval(6,6)[br]}
\put(33,34){\oval(6,6)[tl]}
\put(50,15){\circle*{2}} 
\put(55,5){\line(-1,2){5}}\put(50,15){\vector(1,-2){3}} 
\put(40,15){\circle*{2}} 
\multiput(30,14.5)(0,1){2}{\line(1,0){20}} 
\put(39,40){$\pi$}\put(49,40){$p$}\put(43,9){$D$} 
\thicklines
 \multiput(40,15)(0,2){10}{\line(0,1){1}} 
 \put(50,15){\line(0,1){19}}\put(50,34){\vector(0,-1){12}} 
\thinlines
\end{picture} \qquad
%
%
\begin{picture}(60,45)
\put(10,15){\circle*{2}} 
\put(5,5){\line(1,2){5}}\put(5,5){\vector(1,2){3}} 
\put(20,15){\circle*{2}} 
\multiput(10,14.5)(0,1){2}{\line(1,0){20}} 
\put(19,40){$\pi$}\put(9,40){$p$}\put(13,9){$D$} 
\thicklines
 \multiput(20,15)(0,2){10}{\line(0,1){1}} 
 \put(10,15){\line(0,1){19}}\put(10,15){\vector(0,1){12}} 
\thinlines
\put(30,4){\line(0,1){30}}
\put(27,4){\oval(6,6)[br]}
\put(33,34){\oval(6,6)[tl]}
\put(50,15){\circle*{2}} 
\put(55,5){\line(-1,2){5}}\put(50,15){\vector(1,-2){3}} 
\put(40,15){\circle*{2}} 
\multiput(30,14.5)(0,1){2}{\line(1,0){10}} 
\put(40,15){\line(1,0){10}} 
\put(39,40){$p$}\put(49,40){$\pi$}\put(43,9){$q$} 
\thicklines
 \multiput(50,15)(0,2){10}{\line(0,1){1}} 
 \put(40,15){\line(0,1){19}}\put(40,34){\vector(0,-1){12}} 
\thinlines
\end{picture} \qquad
%
%
\begin{picture}(60,45)
\put(10,15){\circle*{2}} 
\put(5,5){\line(1,2){5}}\put(5,5){\vector(1,2){3}} 
\multiput(10,14.5)(0,1){2}{\line(1,0){20}} 
\put(10,29){\circle*{2}} 
\put(2,40){$p$}\put(18,40){$\pi$} 
\put(1,8){$q$}\put(3,22){${\cal R}$}\put(18,9){$D$} 
\thicklines
 \put(10,29){\line(-1,1){8}}\put(10,29){\vector(-1,1){5}} 
 \multiput(10.0,29.0)(1.4,1.4){7}{\circle*{0.6}} 
 \multiput(10.2,29.2)(1.4,1.4){7}{\circle*{0.6}} 
 \multiput(10.4,29.4)(1.4,1.4){7}{\circle*{0.6}} 
 \put(9.5,15){\line(0,1){14}}              
 \multiput(10.5,15)(0,2){8}{\line(0,1){1}} 
\thinlines
\put(30,4){\line(0,1){30}}
\put(27,4){\oval(6,6)[br]}
\put(33,34){\oval(6,6)[tl]}
\put(50,15){\circle*{2}} 
\put(55,5){\line(-1,2){5}}\put(50,15){\vector(1,-2){3}} 
\put(40,15){\circle*{2}} 
\multiput(30,14.5)(0,1){2}{\line(1,0){10}} 
\put(40,15){\line(1,0){10}} 
\put(39,40){$p$}\put(49,40){$\pi$}\put(43,9){$q$} 
\thicklines
 \multiput(50,15)(0,2){10}{\line(0,1){1}} 
 \put(40,15){\line(0,1){19}}\put(40,34){\vector(0,-1){12}} 
\thinlines
\end{picture}\\[2mm]
$a$ \hspace{45mm} $b$ \hspace{45mm} $c$
\end{center}
\caption{\label{fig4}Interference diagrams for the same process.}
\end{figure}
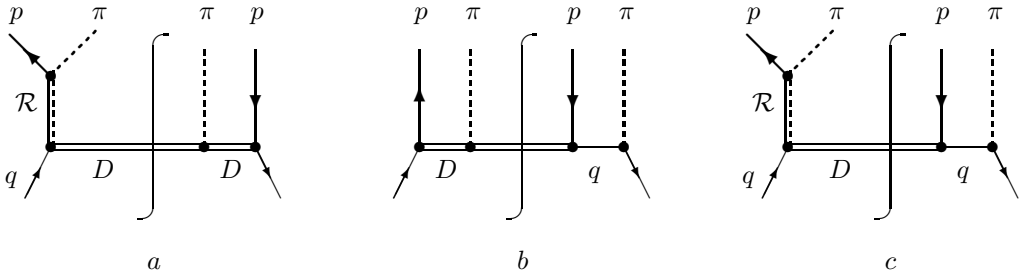

The basic idea of the spectator model is to make a specific ansatz for the
spectral decomposition of the quark correlator by replacing the sum over 
the complete set of intermediate states in Eq.~(\ref{eq:defDelta}) with an 
effective spectator state with a definite 
mass $M_D$ and the quantum 
numbers of the diquark. Consequently, the correlator simplifies to 
\begin{eqnarray} 
\Delta_{ij}(k;P_p,P_\pi)&=& \sumint \; \int
\frac{d^{\,4\!}\zeta}{(2\pi)^4} \; e^{ik\cdot\zeta}\;
\langle 0|\psi_i(\zeta)|\pi,p,X\rangle
\langle X,p,\pi|\ol{\psi}_j(0)|0\rangle \nn\\
&\approx&
\frac{\theta\!\left((k-P_h)^+\right)}{(2\pi)^3} \; 
\delta\left((k-P_h)^2-M_D^2\right)\;
\langle 0|\psi_i(0)|\pi,p,D\rangle\langle D,p,\pi|\ol{\psi}_j(0)|0\rangle 
\nn\\
&\equiv&
\widetilde\Delta_{ij}(k;P_p,P_\pi)\;\delta(\tau_h-\sig_h+M_h^2-M_D^2) \;,
\label{eq:specDelta}
\end{eqnarray}
where in the last line use of the definition (\ref{explinv2}) for the
kinematical invariants has been made. When inserting 
Eq.~(\ref{eq:specDelta}) into 
Eqs.~(\ref{eq:projg-})-(\ref{eq:projsigi-g5}), the additional $\delta$ 
function allows for a drastic simplification of the Dirac 
projections (\ref{eq:projDelta2}), namely
\begin{equation} 
\Delta^{[\Gamma]}(z_h,\xi,{\vec k}_\sT^{\,2},\vec R_\sT^{\, 2},
{\vec k}_\sT \cdot {\vec R}_\sT)=
\left.\frac{\mbox{Tr}[\widetilde\Delta \, \Gamma]}{8(1-z_h)P_h^-}
\right|_{\tau_h=\tau_h(z_h,{\vec k}_\sT^{\,2})} \; ,
\label{eq:projspect}
\end{equation} 
with
\begin{equation}
\tau_h(z_h,{\vec k}_\sT^{\,2})=\frac{z_h}{1-z_h}{\vec k}_\sT^{\,2}
+\frac{M_D^2}{1-z_h}+\frac{M_h^2}{z_h} \;.
\label{eq:tauspect}
\end{equation} 
With this hypothesis, the quark decay described in Fig.~\ref{Delta-2h} is
specialized to the set of diagrams shown in Figs.~\ref{fig3}, \ref{fig4} 
and their hermitean conjugates, where the interference, necessary to 
produce the ``T-odd'' FF, takes place between the channel for direct 
production from the quark $q$ of the proton-pion pair $(p,\pi)$ and the 
channel for the decay of the Roper resonance ${\cal R}$.

By modelling the various ingredients entering these diagrams, we can 
directly calculate the soft hadronic matrix elements of Eq. 
(\ref{eq:specDelta}) and, consequently, make quantitative predictions for 
various Dirac projections and the related FF.

In the following, the most naive picture of the quark structure of the 
nucleon will be assumed, i.e. in the rest frame all quarks are in the 
$1/2^+$ orbitals and the diquark can be in a spin singlet state (scalar 
diquark, indicated by the label $S$) or in a spin triplet state (axial 
vector diquark, indicated by $A$). When generally referring to the 
diquark, the label $D$ will be kept. We will further assume that the 
proton-pion pair has an invariant mass equal to the Roper resonance one, 
so that the diagrams containing an intermediate Roper ${\cal R}$
(Figs.~\ref{fig3}$a$,\ref{fig4}$a$,\ref{fig4}$c$) will be dominant and all 
other channels (including diagrams~\ref{fig3}$b$,\ref{fig3}$c$,
\ref{fig4}$b$ and other resonances) will be neglected.

\subsection{Propagators}
\label{sec:three-a}

Here in the following, we list the propagators needed to compute the 
diagrams \ref{fig3}$a$ and \ref{fig4}$a$,\ref{fig4}$c$.
\begin{itemize}
\item quark with momentum $k$ \\[-6mm]
\begin{center}
\unitlength=1mm
\begin{picture}(30,10)
\thicklines
\put(0,5){\line(1,0){30}}\put(0,5){\vector(1,0){15}}
\put(10,0){$k$}
\thinlines
\end{picture} \hspace{20mm}
      \begin{minipage}{30mm}
      \begin{eqnarray*}
       &&\left(\frac{i}{\kslash -m}\right)_{ij} 
      \end{eqnarray*}
      \end{minipage}
\end{center} 
      
\item Roper with momentum $P_h$ \\[-6mm]
\begin{center}
\unitlength=1mm
\begin{picture}(30,10)
\thicklines
\put(0,5.5){\line(1,0){30}}\multiput(0,4.5)(2,0){15}{\line(1,0){1}}
\put(8,0){$P_h$}\put(14,4.2){$\rangle$}
\thinlines
\end{picture} \hspace{20mm}
      \begin{minipage}{30mm}
      \begin{eqnarray*} 
       &&\left(\frac{i\,(\Pslash_h +M_{\cal R})}{P_h^2-M_{\cal R}^2+
       iM_{\cal R}\Gamma_{\cal R}} \right)_{ij}
      \end{eqnarray*}
      \end{minipage}\end{center} \vspace{-6mm} \noindent
      as it is quoted by the Particle Data Group (PDG)~\cite{pdg} for 
      the determination of the mass and the width of the Roper resonance.
      
\item (scalar/axial vector) diquark with momentum $P_D=P_S/P_A$ \\[-6mm]
\begin{center}
\unitlength=1mm
\begin{picture}(30,10)
\thicklines
\multiput(0,4.5)(0,1){2}{\line(1,0){30}}
\put(8,0){$P_D$}\put(14,4.2){$\rangle$}
\thinlines
\end{picture} \hspace{20mm}
      \begin{minipage}{30mm}
      \begin{eqnarray*} 
        &&\frac{i}{P_S^2-M_S^2}\\[3mm]
        &&\frac{i}{P_A^2-M_A^2}\left(-g_{\mu\nu}+\frac{P_{A\mu}\,P_{A\nu}}
         {M_A^2}\right)
      \end{eqnarray*} 
      \end{minipage}\end{center} \vspace{-6mm} \noindent
      and the polarization sum for the axial vector diquark in the form
      $\displaystyle{\sum_{\lambda}} 
      \epsilon^{*\,\left(\lambda\right)}_{\mu} \ 
      \epsilon^{\left(\lambda\right)}_{\nu} \  = -g_{\mu \nu} + 
      \frac{P_{A\mu}\,P_{A\nu}}{M_A^2}$.
\end{itemize}

\subsection{Vertices}
\label{sec:three-b}

Here in the following, we list and discuss the interaction vertices
$\Upsilon_{ij}$ needed to compute diagrams \ref{fig3}$a$ and
\ref{fig4}$a$,\ref{fig4}$c$. The normalization coefficients have proper
mass dimensions such that 
$\displaystyle{\int d^{\,2\!}\vec k_{\sT}\int d^{\,2\!}\vec R_{\sT}}\;
D_1(z_h,\xi,\vec k_{\sT}^{\, 2},\vec R_{\sT}^{\, 2}, \vec k_{\sT} \cdot 
\vec R_{\sT} )$  be a dimensionless number to be
interpreted as the probability for the hadron pair to carry a $z_h$
fraction of the jet momentum and to share it in $\xi$ and $1-\xi$ parts.

All vertices connecting hadron lines to parton lines will show a form factor
that takes into account the composite structure of the hadron. 
A comment on the arguments of these form factors should be made at this
point. Generally, a form factor can depend on all invariants which can be
built from the momenta of the attached lines. In analogy with previous 
works on the 
spectator model~\cite{melni,jak97} we have chosen a form depending 
on one invariant only, generically denoted $\kappa^2$ in the following. It  
is chosen as the virtuality of the external quark line $\tau_h$ whenever 
the corresponding vertex is attached to it, or as the virtuality of the 
intermediate diquark (quark) propagator for the $A\pi A$-vertex 
($qDp$-vertex) in Fig.~\ref{fig4} diagram $a$ ($c$). 
With these choices of arguments of the form factors, all virtual parton 
lines in the calculation receive an appropriate suppression 
preventing them from being far off-shell. The expressions of form factors 
are such that the net result is an asymptotic behaviour in agreement 
with expectations based on dimensional
counting rules.

\begin{itemize}
\item Roper-proton-pion $({\cal R}p\pi)$ \\[-6mm]
%
%
\unitlength=1mm
\begin{center} 
\begin{picture}(25,35)
\put(10,19){\circle*{2}} 
\put(2,30){$p$}\put(18,30){$\pi$} 
\put(3,12){${\cal R}$}
\thicklines
 \put(10,19){\line(-1,1){8}}\put(10,19){\vector(-1,1){5}} 
 \multiput(10.0,19.0)(1.4,1.4){7}{\circle*{0.6}} 
 \multiput(10.2,19.2)(1.4,1.4){7}{\circle*{0.6}} 
 \multiput(10.4,19.4)(1.4,1.4){7}{\circle*{0.6}} 
 \put(9.5,8){\line(0,1){11}}              
 \multiput(10.5,8)(0,2){6}{\line(0,1){1}} 
\thinlines
\end{picture}     \hspace{15mm} 
      \begin{minipage}[b]{60mm}
      \begin{eqnarray*}
       & &\Upsilon^{{\cal R}p\pi}_{ij}=f_{{\cal R}p\pi} \;  [\g_5 ]_{ij} \; 
       \equiv g \;  [\g_5 ]_{ij} \; ,
      \end{eqnarray*}
      \vspace{1mm}
      \end{minipage}
\end{center} \vspace{-6mm} \noindent 
      where $g^2/4 \pi=14.3$ is the strong coupling constant of the 
      $\pi NN$ pseudoscalar interaction. Calculation of the 
      Roper-proton-pion coupling $f_{{\cal R}p\pi}$ by means of two 
      independent ways, namely the Roper decay and the elastic 
      $p+\pi\rightarrow p+\pi$ scattering through a virtual channel with 
      the same mass and width of the Roper, produces a decay width and an 
      empirical channel strength, respectively, that depend on 
      $f_{{\cal R}p\pi}$. Comparison with experimental data allows 
      $f_{{\cal R}p\pi}$ to be compatible with $g$ within the large 
      experimental error bars. Hence, in the 
      following in all vertices involving a Roper we will keep the same 
      form and strength as in the ones involving a proton, also because, 
      being the quark content the same, the asymptotic behaviour of the 
      form factors should be the same.

\item quark-diquark-(Roper/proton) $(qD{\cal R}/qDp)$ \\[1mm]
\begin{center} \begin{minipage}{25mm}
\unitlength=1mm
%
%
\begin{picture}(25,25)
\put(10,15){\circle*{2}} 
\put(5,5){\line(1,2){5}}\put(5,5){\vector(1,2){3}} 
\multiput(10,14.5)(0,1){2}{\line(1,0){10}} 
\put(1,8){$q$}\put(3,22){${\cal R}$}\put(18,9){$D$} 
\thicklines
 \put(9.5,15){\line(0,1){11}}              
 \multiput(10.5,15)(0,2){6}{\line(0,1){1}} 
\thinlines
\end{picture}\\
%
%
\begin{picture}(25,25)
\put(10,15){\circle*{2}} 
\put(5,5){\line(1,2){5}}\put(5,5){\vector(1,2){3}} 
\multiput(10,14.5)(0,1){2}{\line(1,0){10}} 
\put(1,8){$q$}\put(3,22){$p$}\put(18,9){$D$} 
\thicklines
\put(10,15){\line(0,1){11}}\put(10,15){\vector(0,1){7}} 
\thinlines
\end{picture}
      \end{minipage} \hspace*{10mm}
      \begin{minipage}{110mm}
      \begin{eqnarray*} 
      \Upsilon^{qS{\cal R}/qSp}_{ij}&=& f_{(qS{\cal R}/qSp)}(\kappa^2) \; 
      \one_{ij} \equiv N_{qS} \; 
      \frac{\kappa^2-m^2}{|\kappa^2 - \Lambda^2|^{\alpha}} \; 
      \one_{ij} \\[8mm]
      \Upsilon_{ij}^{qA{\cal R}/qAp,\ \mu}&=&f_{(qA{\cal R}/qAp)}(\kappa^2)\; 
      \left[\g_5\g^\mu\right]_{ij} \equiv \frac{N_{qA}}{\sqrt{3}}\; 
      \frac{\kappa^2 -m^2}{|\kappa^2 -\Lambda^2|^{\alpha}}\; 
      \left[\g_5\g^\mu\right]_{ij} \; ,
      \end{eqnarray*} 
      \end{minipage} \end{center} \vspace*{-6mm}\noindent
      where $m$ is the quark mass and $\Lambda=0.5$ GeV is a cut-off 
      parameter that excludes large virtualities of the quark. This choice 
      of the form factor has the advantage of killing the pole of the 
      quark propagator~\cite{melni,jak97}. As remarked previously the argument
      of the (qDp) form factor is $\kappa^2=\tau_h$ in diagram \ref{fig4}$(a)$
      and $\kappa^2=(k-P_\pi)^2$ in diagram \ref{fig4}$(c)$. The power 
      law $\alpha$ is 
      determined consistently with the quark counting rule that drives the 
      asymptotic behaviour of the FF at large $z_h$~\cite{joflipat}, i.e. 
      \begin{equation}
      (1-z_h)^{2 \alpha -1} \equiv (1-z_h)^{-3+2r+2|\lambda|} \; ,
      \label{eq:qcr}
      \end{equation}
      where $r$ is the number of constituent quarks in the considered hadron, 
      and $\lambda$ is the difference between the quark and hadron helicities. 
      Thus, in this case $\alpha=2$. To determine the form of the vertex
      involving the axial vector diquark we took inspiration 
      from considerations which are valid for on-shell particles, only.
      In that case, the most general form for the vertex
      involving the axial vector diquark is $\g_5 \g^\mu + \g_5 (a p^\mu + 
      b k^\mu)$, with $a,b$ arbitrary 
      parameters. The second piece can be rewritten using the
      difference $p^\mu - k^\mu = P_A^\mu$ and the sum $(p+k)^\mu$ of the
      momenta involved. The former is ruled out because in configuration
      space it is equivalent to taking the 4-divergence of the spin-1
      diquark field, $\partial_{\mu} \psi_A^{\mu}$: the Lorentz condition 
      $\partial_{\mu} \psi_A^{\mu} = 0$ is usually taken to grant a field
      energy which is positive defined. The latter can be transformed, 
      using a Gordon-like decomposition, into the linear combination of 
      $\g_5 \g^\mu$ (to be reabsorbed into the initial first piece) and 
      $\g_5 \sigma^{\mu \nu} P_{A \nu}$ which, contrary to
      the axial current, is even under $G$-parity transformations and,
      therefore, is forbidden~\cite{okun}. Hence, the general structure 
      $\g_5 \g^{\mu}$ is assumed for this vertex.
      The overall 
      normalizations $N_{qS}=7.92$ GeV$^2$ and $N_{qA}=11.557$ GeV$^2$ 
      are fixed by computing the second moment of $D_1(z_h)$ and 
      comparing it with available data; their values, as well as the 
      $\Lambda$ parameter, are taken directly from Ref.~\cite{jak97}.

\item quark-pion-quark $(q\pi q)$ \\[-5mm]
\begin{center}
\unitlength=1mm
%
%
\begin{picture}(35,25)
\put(20,5){\circle*{2}} 
\put(10,5){\line(1,0){20}}
\put(10,5){\vector(1,0){4.5}}\put(20,5){\vector(1,0){5.5}} 
\put(19,20){$\pi$}\put(5,4){$q$}\put(32,4){$q$} 
\thicklines
 \multiput(20,5)(0,2){6}{\line(0,1){1}} 
\thinlines
\end{picture} \hspace{10mm}
      \begin{minipage}[b]{80mm}
      \begin{eqnarray*}
      & &\Upsilon^{q\pi q}_{ij}= f_{q\pi q}(\kappa^2) \; [ \gamma_5 ]_{ij} 
      \equiv N_{q\pi} \; \frac{\kappa^2 -m^2}{|\kappa^2
         -\Lambda_{\pi}^2|^{\alpha}} \; [ \gamma_5 ]_{ij} \; ,
      \end{eqnarray*}
      \end{minipage}
\end{center} \vspace{-4mm} \noindent
      where the new parameters $\Lambda_{\pi}=0.4$ GeV, $N_{q\pi}=2.564$
      GeV have the same meaning and role as before 
      and are determined in a similar manner~\cite{jak97}. In this case, Eq.
      (\ref{eq:qcr}) gives $\alpha=3/2$.
      
\vspace{1cm}

\item (axial) diquark-pion-(axial) diquark $(A\pi A)$ \\[-5mm]
\begin{center}
\unitlength=1mm
%
%
\begin{picture}(35,25)
\put(20,5){\circle*{2}} 
\multiput(10,4.5)(0,1){2}{\line(1,0){20}} 
\put(19,20){$\pi$}\put(5,4){$\mu$}\put(32.6,4){$\nu$} 
\put(12,-1){$P_A'$}\put(24,-1){$P_A$} 
\thicklines
\multiput(13,4)(12,0){2}{$\large \rangle$}
 \multiput(20,5)(0,2){6}{\line(0,1){1}} 
\thinlines
\end{picture} \hspace{6mm}
      \begin{minipage}[b]{80mm}
      \begin{eqnarray*}
      & \Upsilon^{A\pi A}_{\mu\nu} &= i \, f_{A\pi A}(\kappa^2) \;  
      \eps_{\mu\nu\rho\sig} \,P_A'^\rho\,P_A^\sig \nonumber \\
      & &= - i \; N_{\pi A} \; \frac{\kappa^2 -M_A^2}{|\kappa^2
         -\Lambda_{\pi}^2|^{\alpha}} \; \eps_{\mu\nu\rho\sig} 
         \,P_A'^\rho\,P_A^\sig \;.
      \end{eqnarray*}
      \end{minipage}
\end{center}  \noindent
      Only the axial diquarks are here considered because two scalar
      diquarks cannot couple to a pseudoscalar particle. The overall
      negative sign, together with the Dirac structure and the strength 
      $N_{\pi A}$, are justified the Appendix. The 
      parameters $\Lambda_{\pi}$ and $\alpha$ are determined as before. 
      In this case, Eq. (\ref{eq:qcr}) gives $\alpha = 2$.
      
\end{itemize}

\subsection{Calculation of fragmentation functions}
\label{sec:three-c}

Using the results of the previous
Sections~\ref{sec:three-a}, \ref{sec:three-b} we can calculate the soft
hadronic matrix elements of Eq.~(\ref{eq:specDelta}) corresponding to
the diagrams~\ref{fig3}$a$,\ref{fig4}$a$,\ref{fig4}$c$ and their hermitean 
conjugates; consequently, all the Dirac projections needed to determine the FF 
$D_1, G_1^{\perp}, H_1^{\perp}, H_1^{\newangle}$ can be computed according to 
Eq.~(\ref{eq:projspect}) and (\ref{eq:projg-})-(\ref{eq:projsigi-g5}).

In agreement with theoretical expectations~\cite{noi}, the 
``T-even'' $D_1$ gets the dominant contribution from the diagonal diagram 
\ref{fig3}$a$, while the naive ``T-odd'' 
$G_1^{\perp}$, $H_1^{\perp}$, $H_1^{\newangle}$ are entirely driven by the
interference diagrams \ref{fig4}$a$,\ref{fig4}$c$, thus confirming that 
in this 
case there is a close relation between the ``T-odd'' structure of the cross
section to leading order and the FSI produced by the interference of the 
two channels for proton-pion production. Therefore, in the following, we 
will concentrate only on the ``T-odd'' FF.

Before addressing the results, some remarks must be added about the flavor
decomposition of FF, in particular about the different role played by the
scalar and the axial diquarks. In fact, as demanded by Pauli principle, if
the total spin-flavor wave function of a baryon must be symmetric, the 
scalar diquark must be in a flavor singlet state while the axial diquark 
in a flavor triplet state. This leads, e.g., to the known $SU(4)$ 
structure of the proton wave function~\cite{jak97} and to the ratio 3:1 
between the contributions of the scalar and axial diquarks, respectively, 
for the fragmentation of an $u$ quark. We assume that the spin-flavor 
content of the Roper wave function is similar to the proton one. 
Therefore, Eqs.~(\ref{eq:projg-g5}), (\ref{eq:projsigi-g5}) for the 
fragmentation of an $u$ quark into a proton and a pion, 
$u \rightarrow p+\pi$, get contributions from 
\begin{eqnarray}
\Delta^{[\g^- \g_5]}(z_h,\xi,{\vec k}_\sT^{\,2},\vec R_\sT^{\, 2},
{\vec k}_\sT \cdot {\vec R}_\sT) &\equiv &
\frac{\eps_\sT^{ij} \,R_{\sT i}\,k_{\sT j}}{M_1\,M_2} \; G_1^\perp = 
   \left.\frac{\mbox{Tr}[\widetilde\Delta \, \g^- \g_5]}{8(1-z_h)P_h^-}
    \right|_{\tau_h=\tau_h(z_h,{\vec k}_\sT^{\,2})} \nn\\ 
&= &\left.\frac{1}{8(1-z_h)P_h^-} \; \mbox{Tr}\left[ \left( 
    \frac{1}{2} \widetilde\Delta_{IaA} + \frac{3}{2} \widetilde\Delta_{IcS} + 
    \frac{1}{2} \widetilde\Delta_{IcA} \right) \g^- \g_5 \right] 
    \right|_{\tau_h=\tau_h(z_h,{\vec k}_\sT^{\,2})} \label{eq:projg-g5tot}
    \\[3mm]
\Delta^{[i\sig^{i-} \g_5]}(z_h,\xi,{\vec k}_\sT^{\,2},\vec R_\sT^{\, 2},
{\vec k}_\sT \cdot {\vec R}_\sT) &\equiv &
{\epsilon_\sT^{ij}R_{\sT j}\over M_1+M_2}\, 
H_1^{\newangle}+{\epsilon_\sT^{ij}k_{\sT j}\over M_1+M_2}\,H_1^\perp = 
   \left.\frac{\mbox{Tr}[\widetilde\Delta \, i \sig^{i-} 
   \g_5]}{8(1-z_h)P_h^-} \right|_{\tau_h=\tau_h(z_h,{\vec k}_\sT^{\,2})} \nn\\
&= &\left.\frac{1}{8(1-z_h)P_h^-} \; \mbox{Tr}\left[ \left( 
    \frac{1}{2} \widetilde\Delta_{IaA} + \frac{3}{2} \widetilde\Delta_{IcS} + 
    \frac{1}{2} \widetilde\Delta_{IcA} \right) i \sig^{i-} \g_5 \right] 
    \right|_{\tau_h=\tau_h(z_h,{\vec k}_\sT^{\,2})} \label{eq:projsigi-g5tot} 
    \;
    ,
\end{eqnarray}
where $\widetilde\Delta_{I(a/c)(S/A)}$ refers to the soft matrix element of
diagram \ref{fig4}$(a/c)$ involving the (scalar/axial) diquark, 
respectively. Their relative weights are given by the proper Clebsch-Gordan 
coefficients for the combination of isospins according to the above 
mentioned $SU(4)$ symmetry. The latter is responsible also for the 
equivalence $(u \rightarrow p) \Leftrightarrow (d \rightarrow n)$, 
therefore the previous equation can usefully be adopted to describe the $d$ 
quark fragmentation into a neutron and a pion. On similar grounds, the 
fragmentation $u \rightarrow n+\pi$, or equivalently 
$d \rightarrow p+\pi$, gets contribution only from the diagrams
involving the axial diquark only~\cite{jak97}.

By working out the matrix elements included in Eqs.
(\ref{eq:projg-g5tot}), (\ref{eq:projsigi-g5tot}), we get the explicit
expressions for the naive ``T-odd'' FF for the process 
$u \rightarrow p+\pi$ at leading order in the context of the diquark 
spectator model, i.e.
\begin{eqnarray}
\lefteqn{
G_1^{\perp\; u\rightarrow p+\pi}(z_h,\xi,\tau_h,M_h^2,\sig_d)=
\frac{2}{(2\pi)^3}\;
\frac{\Gamma_{\cal R} M_{\cal R}}{(M_h^2-M_{\cal R}^2)^2+M_{\cal
R}^2\Gamma_{\cal R}^2}\;
\frac{M_pM_\pi}{1-z_h}\;\Bigg\{}\nn\\
&&\hspace{4mm} {}+\frac{1}{2}\;
\frac{2 \; f_{A\pi A}\;f_{{\cal R}p\pi}\;f_{qAp}\;f_{qA{\cal R}}}
     {3 (\tau_h-m^2)^2\,\left(\tau_h-M_h^2+2M_p^2-M_A^2-\sig_d\right)} \; 
\bigg\{m\left(M_h^2-2M_pM_{\cal R}+M_p^2-M_\pi^2\right)\bigg\}\nn\\[4mm]
&&\hspace{4mm}{}+\frac{3}{2}\; 
\frac{f_{q\pi q}\;f_{qSp}\;f_{{\cal R}p\pi}\;f_{qS{\cal R}}}
     {\,(\tau_h-m^2)^2\;\left(\tau_h-M_h^2+2M_\pi^2+M_S^2-2m^2+
     \sig_d\right)} \; 
\bigg\{\tau_h-m^2+2\,m(M_{\cal R}-M_p)\bigg\}\nn\\[4mm]
&&\hspace{4mm} {}-\frac{1}{2}\;
\frac{f_{q\pi q}\;f_{qAp}\;f_{{\cal R}p\pi}\;f_{qA{\cal R}}}
     {6 M_A^2 \; (\tau_h-m^2)^2\;\left(\tau_h-M_h^2+2M_\pi^2+M_A^2-2m^2+
     \sig_d\right)}\;
\bigg\{-\left(\tau_h-m^2\right)^2\nn\\
&&\hspace{9mm} {}+\left(\tau_h+m^2+2 m M_{\cal R}\right)
\left[M_h^2+2(M_p^2-M_\pi^2)-M_A^2+m^2+2 m(2 M_p-M_{\cal R})-\sig_d\right]\nn\\
&&\hspace{9mm} -4\,m\,\left[M_p(M_h^2-2 M_A^2+m^2)+m M_h^2
               {}+(2 M_A^2+m M_{\cal R})(2 M_p-M_{\cal R})\right]\bigg\} 
               \hspace{3mm} \Bigg\}
\label{eq:g1perpspect}
\end{eqnarray} 

\begin{eqnarray}
\lefteqn{
H_1^{\newangle \; u \rightarrow p+\pi}(z_h,\xi,\tau_h;M_h^2,\sig_d)=
\frac{2}{(2\pi)^3}\; \frac{\Gamma_{\cal R} M_{\cal R}}{(M_h^2-M_{\cal R}^2)^2+
M_{\cal R}^2\Gamma_{\cal R}^2}\;
\frac{M_p+M_\pi}{z_h\,(1-z_h)}\;\Bigg\{}\nn\\
&&\hspace{4mm} {}+\frac{1}{2}\;
   \frac{f_{A\pi A}\;f_{{\cal R}p\pi}\;f_{qAp}\;f_{qA{\cal R}}\;
      \left(M_h^2+M_p^2-M_\pi^2-2 M_p M_{\cal R}\right)}
     {3 (\tau_h-m^2)^2\;\left(\tau_h-M_h^2+2M_p^2-M_A^2-\sig_d\right)}
     \bigg\{M_h^2-M_A^2-z_h\tau_h+m^2(1-z_h)\bigg\}\nn\\[4mm]
&&\hspace{4mm}{}+\frac{3}{2}\;
\frac{f_{q\pi q}\;f_{qSp}\;f_{{\cal R}p\pi}\;f_{qS{\cal R}}}
     {2 (\tau_h-m^2)^2 \; \left(\tau_h-M_h^2+2M_\pi^2+M_S^2-2m^2+
     \sig_d\right)}\nn\\
&&\hspace{12mm}\times\bigg\{2[M_h^2-M_S^2+m^2(1-2z_h)](M_{\cal R}-M_p)
   +2(\tau_h-m^2)[(1-z_h)(M_{\cal R}-M_p)-z_h m-M_{\cal R}]\bigg\}\nn\\[4mm]
&&\hspace{4mm}{}-\frac{1}{2} \; 
\frac{f_{q\pi q}\;f_{qAp}\;f_{{\cal R}p\pi}\;f_{qA{\cal R}}}
     {6 M_A^2 \; (\tau_h-m^2)^2\;\left(\tau_h-M_h^2+2M_\pi^2+M_A^2-2m^2+
     \sig_d\right)}\nn\\
&&\hspace{12mm}\times\bigg\{\tau_h^2 \left(-2 z_h M_p+z_h M_{\cal R}-m\right)
   +M_h^2 [M_h^2+(1-z_h)m^2](M_{\cal R}-2 M_p-m)-m^3 M_h^2(1-z_h)\nn\\
&&\hspace{15mm}{}+(M_p^2-M_\pi^2)[2 (1-z_h) m^2 M_{\cal R}+2 (M_h^2-M_A^2) 
   (M_{\cal R}+m)]+M_A^4 (4 M_p-3 M_{\cal R}+m)\nn\\
&&\hspace{15mm}{}+ M_A^2 \left[-2 M_h^2 (M_p-M_{\cal R})+m^2 [-2 (1-2 z_h) M_p
   +3 (1-z_h) M_{\cal R}-2 m]\right]\nn\\
&&\hspace{15mm}{}+\tau_h\bigg[
   2(M_p^2-M_\pi^2)[m (1-2 z_h)-z_h M_{\cal R}]+\sig_d[-m (1-2 z_h)+z_h M_{\cal
   R}]+M_h^2 (1+z_h) (2 M_p-M_{\cal R})\nn\\
&&\hspace{23mm}{}+M_A^2 [M_{\cal R} (1-3 z_h)-4 M_p (1-z_h)+2 z_h m]
   +m [2 m^2 (1-z_h)+m (1-z_h) (2 M_p-M_{\cal R})+2 M_h^2]\bigg] \nn\\
&&\hspace{15mm}{}+\sig_d [(M_A^2-M_h^2) (M_{\cal R}+m)-m^2 M_{\cal R} (1-z_h)]
   \bigg\}\hspace{3mm}\Bigg\}
\label{eq:h1anglespect}
\end{eqnarray} 

\begin{eqnarray}
\lefteqn{
H_1^{\perp \; u\rightarrow p+\pi}(z_h,\xi,\tau_h,M_h^2,\sig_d)=
\frac{2}{(2\pi)^3}\;
\frac{\Gamma_{\cal R} M_{\cal R}}{(M_h^2-M_{\cal R}^2)^2+M_{\cal
R}^2\Gamma_{\cal R}^2}\; \frac{M_p+M_\pi}{1-z_h}\;\Bigg\{}\nn\\
&&\hspace{4mm} {}+\frac{1}{2}\;
\frac{f_{A\pi A}\;f_{{\cal R}p\pi}\;f_{qAp}\;f_{qA{\cal R}}\;
\left(M_h^2+M_p^2-M_\pi^2-2 M_p M_{\cal R}\right)}
     {6 (\tau_h-m^2)^2\;\left(\tau_h-M_h^2+2M_p^2-M_A^2-\sig_d\right)}\nn\\
&&\hspace{12mm}\times
\bigg\{-(\tau_h-m^2)+\sig_d+(M_h^2-M_A^2+m^2)(1-2\xi) \bigg\}\nn\\[4mm]
&&\hspace{4mm} {}+\frac{3}{2} \; 
\frac{f_{q\pi q}\;f_{qSp}\;f_{{\cal R}p\pi}\;f_{qS{\cal R}}}
     {2 (\tau_h-m^2)^2\;\left(\tau_h-M_h^2+2M_\pi^2+M_S^2-2m^2+
     \sig_d\right)}\nn\\
&&\hspace{12mm}\times\bigg\{
    (1-2\xi)\left(M_h^2-M_S^2+m^2\right)(M_{\cal R}-M_p)+
    \left(\sig_d+\tau_h-m^2\right)(M_{\cal R}-M_p)
    -2(1-\xi)\left(\tau_h-m^2\right)M_p\bigg\}\nn\\[4mm]
&&\hspace{4mm} {}-\frac{1}{2} \; 
\frac{f_{q\pi q}\;f_{qAp}\;f_{{\cal R}p\pi}\;f_{qA{\cal R}}}
     {12 M_A^2 \; (\tau_h-m^2)^2\;\left(\tau_h-M_h^2+2M_\pi^2+M_A^2-2m^2+
     \sig_d\right)}\nn\\
&&\hspace{12mm}\times
    \bigg\{-\tau_h^2 \left[m (1-2\xi)-(2 M_p-M_{\cal R})\right] -\sig_d^2 
    (m+M_{\cal R})+2 M_A^2 m^2 \left[2 M_p (1+\xi)-3\xi M_{\cal R}-m (1-2\xi)
    \right]\nn\\
&&\hspace{15mm}{}+(M_p^2-M_\pi^2)
    \left[2 (M_h^2-M_A^2)(m+M_{\cal R})(1-2\xi)-4 m^2 (\xi M_{\cal R}+m)
    \right]\nn\\
&&\hspace{15mm}{}-(M_h^4-M_A^4) (2 M_p-M_{\cal R}+m) (1-2\xi)
    +2 M_A^2 (M_A^2-M_h^2) (M_p-M_{\cal R}) (1-2\xi) \nn\\
&&\hspace{15mm}{}+2 M_h^2 m^2 \left[\xi (2 M_p-M_{\cal R})-m (1-2\xi)
   \right]\nn\\
&&\hspace{15mm}{}+\tau_h \bigg[2 \sig_d [M_p-M_{\cal R}-m (1-\xi)]
    +2 M_A^2 [M_{\cal R} (2-\xi)-M_p (5-4\xi)]+2\xi M_{\cal R} (m^2+M_h^2)
    \nn\\
&&\hspace{23mm}{}+(M_p^2-M_\pi^2) [m (6-4\xi)+2 M_{\cal R}]
    +m [2 (M_h^2+m^2) (1-2\xi)-4 m\xi M_p]-4\xi M_p M_h^2\bigg]\nn\\
&&\hspace{15mm}
  -2 \sig_d \bigg[M_A^2 [2 (M_p-M_{\cal R})+\xi (M_{\cal R}+m)]
         +m [(1-\xi) M_h^2-m (m+\xi M_{\cal R})]-M_h^2 (\xi M_{\cal R}-M_p)
         \nn\\
&&\hspace{24mm}
  -(M_{\cal R}+m) (M_p^2-M_\pi^2)\bigg]\bigg\} \hspace{3mm} \Bigg\} \; ,
\label{eq:h1perpspect}
\end{eqnarray} 
where $M_p,M_{\pi},M_S,M_A$ are the proton, pion, scalar and axial diquark 
masses, respectively, and $M_{\cal R},\Gamma_{\cal R}$ are the mass and the 
experimental width of the Roper
resonance. The explicit dependence of FF upon the invariants described in 
Eq.~(\ref{explinv2}), rather than on the variables shown in 
Eqs.~(\ref{eq:projg-g5tot}), (\ref{eq:projsigi-g5tot}), is due just 
to convenience of 
making the formula the most easily readable. In the following, plots will 
be shown of FF considered again as functions of 
$z_h,\xi,{\vec k}_\sT^{\,2},{\vec R}_\sT^{\,2},{\vec k}_\sT\cdot{\vec R}_\sT$,
 which allow for a better 
understanding of the underlying physics, as stressed in 
Sec.~\ref{sec:two}.

\section{Numerical Results}
\label{sec:four}

Plots of FF in Eqs.~(\ref{eq:g1perpspect})-(\ref{eq:h1perpspect}) are 
shown for the parameter values (in GeV)
\begin{equation}
m=0.36\;, \; M_S=0.6\; , \; M_A=0.8\; , \; M_p=0.9383\; , \; 
M_{\pi}=0.1396 \; .
\label{eq:masses}
\end{equation}
As already anticipated in Sec.~\ref{sec:three}, the proton-pion invariant 
mass will be taken equal to the Roper resonance mass, $M_h=M_{\cal R}=1.44$ 
GeV, so that diagrams of Figs.~\ref{fig3}$b$, \ref{fig3}$c$, \ref{fig4}$b$ are 
then negligible. The Roper resonance width is $\Gamma_{\cal R}=0.35$ 
GeV~\cite{pdg}. 

For the purpose of displaying our results 
we will choose a special kinematics where 
${\vec k}_\sT\cdot{\vec R}_\sT=0$, namely where the hadron-pair plane is 
perpendicular to the plane containing the jet axis and the direction of the 
hadron pair $\vec P_h$ (see Fig.~\ref{fig2}, where the indicated 
angle $\alpha$ 
takes the value $90^{\rm o}$). Consequently, it turns 
out from Eq. (\ref{eq:sigd2}) that $\sig_d=\sig_d(z_h,\xi,{\vec k}_\sT^{\,2},
{\vec R}_\sT^{\,2})$. Moreover, from Eq. (\ref{eq:tauspect}) also the 
invariant $\tau_h$ depends on $z_h,{\vec k}_\sT^{\,2}$ and from Eq. 
(\ref{eq:pt2}) we have ${\vec R}_\sT^{\,2} (\xi)$. Therefore, at fixed 
invariant mass $M_h=M_{\cal R}$ and ${\vec k}_\sT\cdot{\vec R}_\sT=0$ the FF 
are actual functions of $z_h,\xi,{\vec k}_\sT^{\,2}$ only. 
Equation~(\ref{eq:pt2}) 
further constrains the variable $\xi$, because from the positivity of 
${\vec R}_\sT^{\,2} (\xi)$ we have
\begin{equation}
M_h^2 \ge \frac{M_p^2}{\xi}+\frac{M_{\pi}^2}{1-\xi} \; ,
\label{eq:xilimit}
\end{equation}
and, for the above fixed values of the masses, $0.43 \le \xi \le 0.98$. 

In Fig.~\ref{fig5} the interference FF 
$H_1^{\newangle \; u\rightarrow p+\pi}$, 
$H_1^{\perp \; u\rightarrow p+\pi}$, 
$G_1^{\perp \; u\rightarrow p+\pi}$ are shown 
(from top to bottom, respectively) as functions of $z_h$, 
${\vec k}_\sT^{\,2}$ at $\xi=0.7$. Similar results are obtained for 
different values of $\xi$ in the
allowed range. The maximum sensitivity to the fragmentation mechanism is 
concentrated around the kinematical range where the pair takes roughly 80\% of 
the jet energy ($z_h \sim 0.8$) and has a small 
transverse momentum 
with respect to the jet axis ($\vec k_\sT^2 \lesssim 0.4$ GeV$^2$; we 
recall that $G_1^{\perp}$, $H_1^{\newangle}$, $H_1^{\perp}$ are defined 
as the probability 
difference for the unpolarized hadron pair to be generated from the 
fragmentation of a quark with parallel or transverse polarization with 
respect to its longitudinal momentum). 

It is worth noting also from the plot scales that $G_1^{\perp}$ is roughly
smaller by one order of magnitude with respect to the other pair of FF.

\begin{minipage}{19cm}
\begin{center}
\hspace*{-2cm}
\epsfig{file=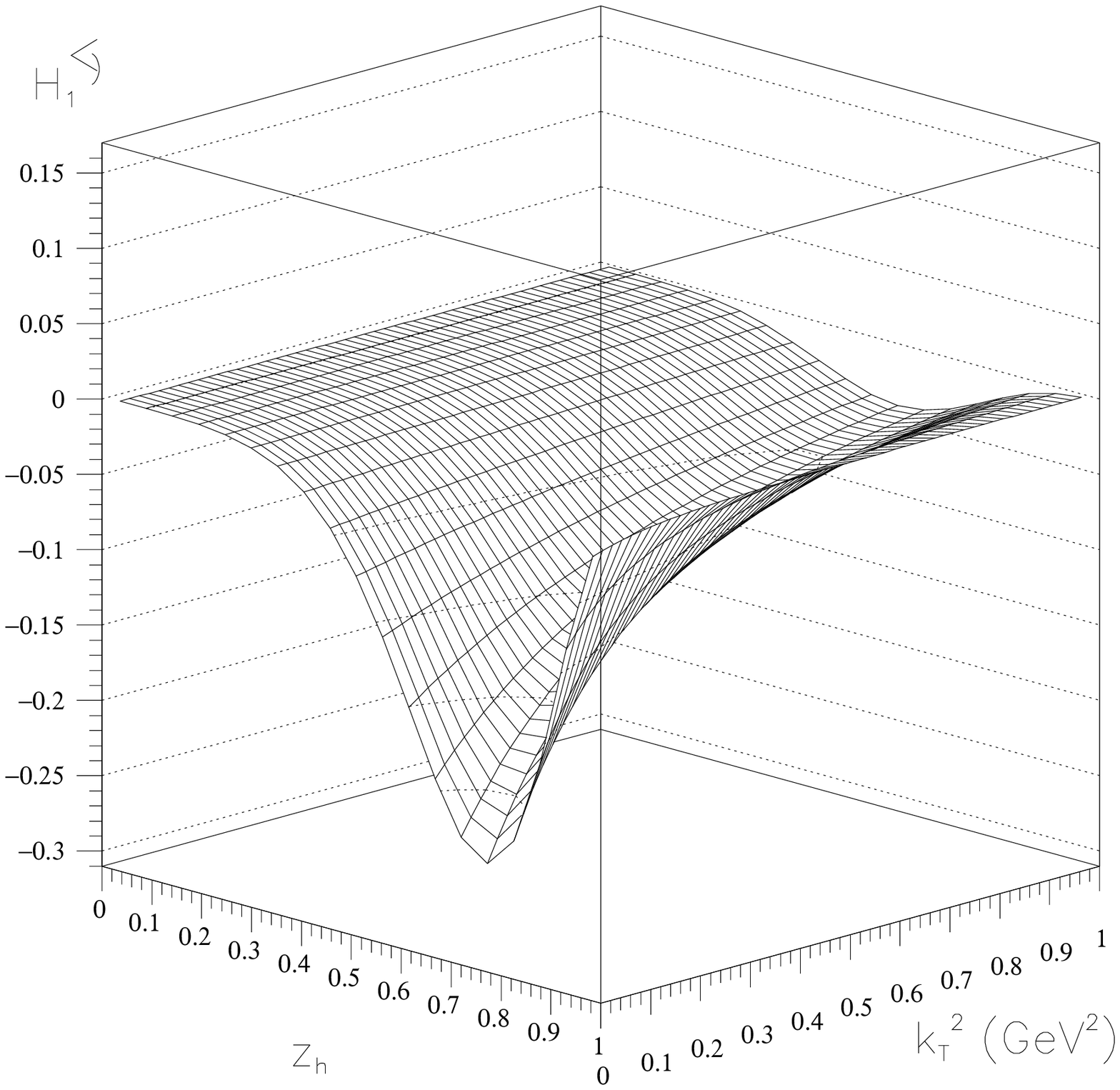, width=9cm} 
\hspace{-.5cm}  
\epsfig{file=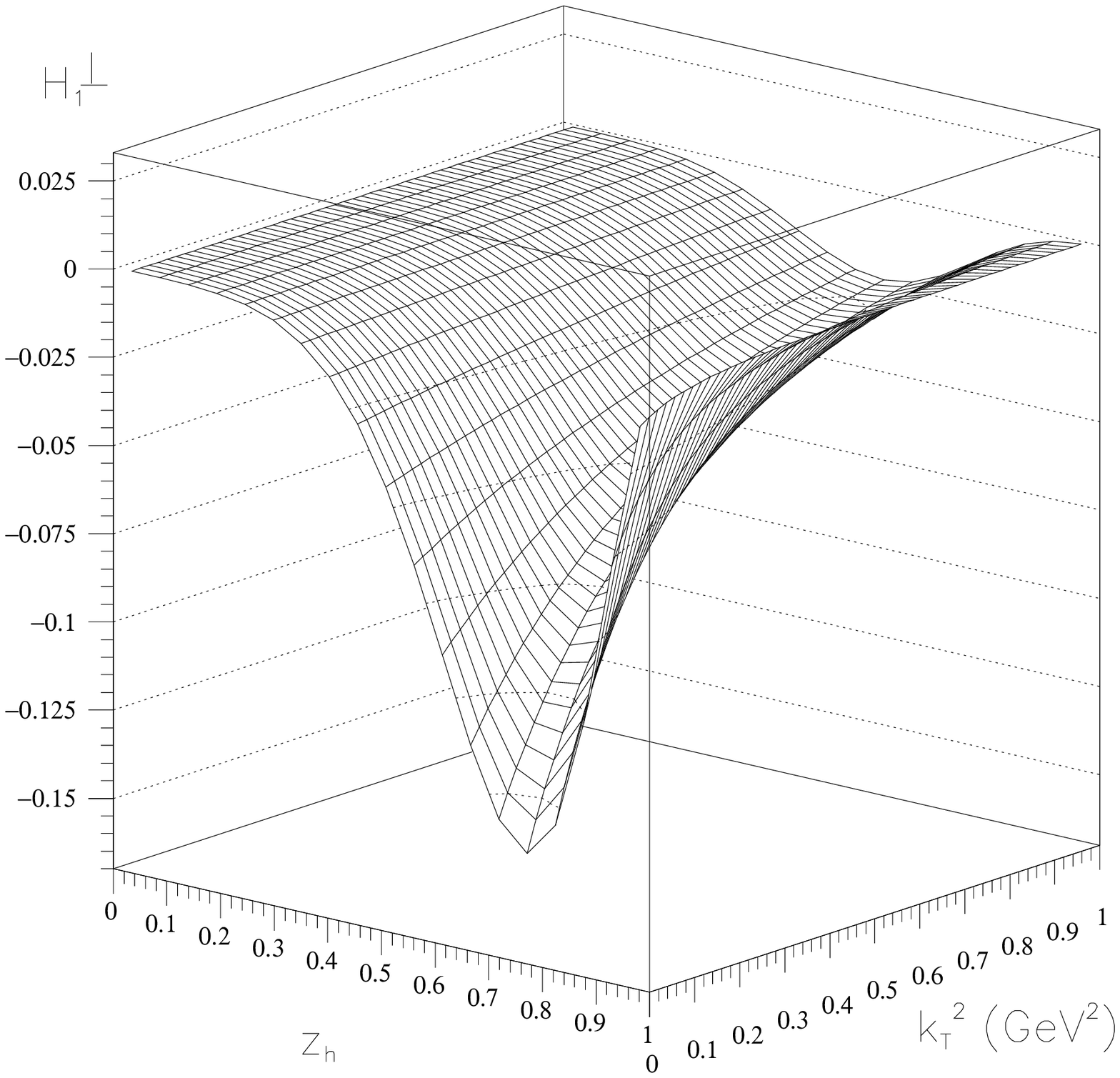, width=9cm} 
\hfil \\
\vspace{-.8cm}
\hspace*{-1cm}
\epsfig{file=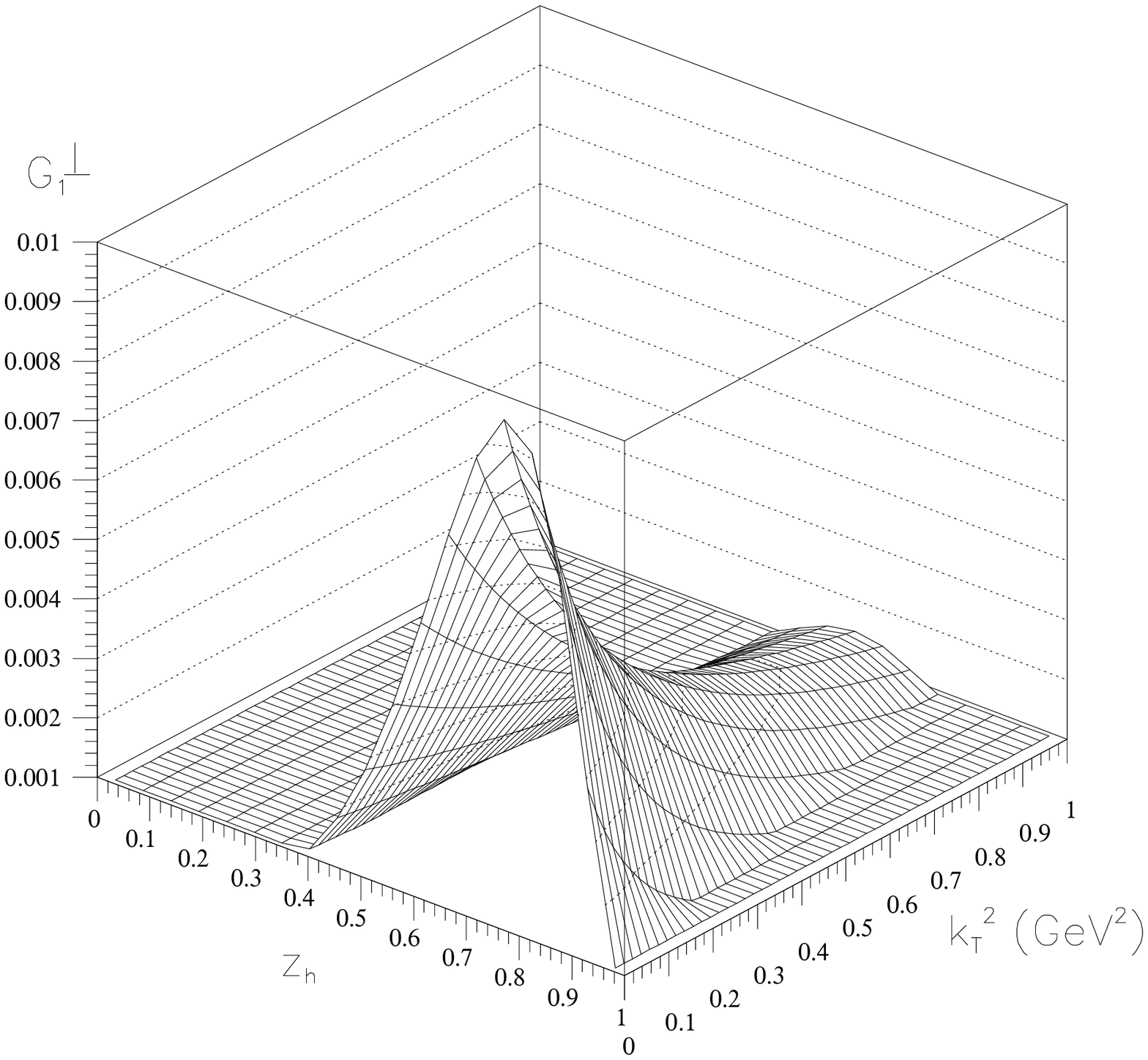, width=9cm}
\end{center}
\end{minipage}
\begin{figure}[h]
\caption{\label{fig5} $H_1^{\newangle}(z_h,{\vec k}_\sT^{\,2})$, 
$H_1^{\perp}(z_h,{\vec k}_\sT^{\,2})$, 
$G_1^{\perp}(z_h,{\vec k}_\sT^{\,2})$ at 
$\xi=0.7$ (from top to bottom) for the fragmentation of a quark 
$u$ into a proton $p$ and a pion $\pi$. Kinematics is chosen such that the 
invariant mass of the pair is equal to the Roper resonance and 
${\vec k}_\sT\cdot{\vec R}_\sT=0$ (see Fig.~\protect{\ref{fig2}}).}
\end{figure}
 
\begin{figure}[htb]
\begin{center}
\psfig{file=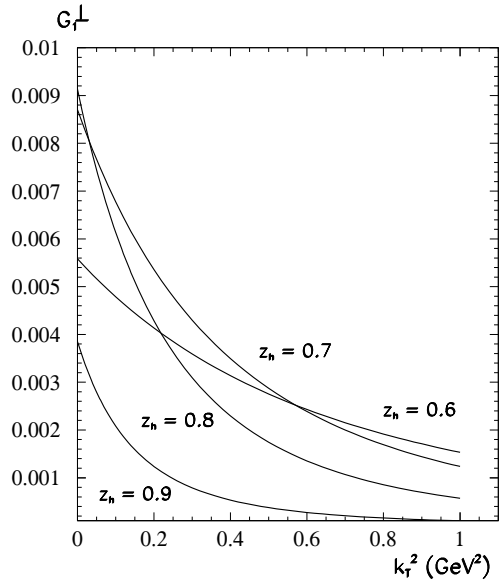, width=7cm}
\end{center}
\caption{\label{fig6} $G_1^{\perp}({\vec k}_\sT^{\,2})$ at 
$\xi=0.7$ and $z_h=0.6,0.7,0.8,0.9$ for the fragmentation of a quark 
$u$ into a proton $p$ and a pion $\pi$. Kinematics as in previous 
Fig.~\protect{\ref{fig5}}.}
\end{figure}

By ``cutting'' the 3-d surfaces of 
Fig.~\ref{fig5} at constant values of $z_h$ in the interesting 
range, e.g.\ for $z_h \ge 0.6$, one can get the trend of the dependence on
transverse quark momenta in the non-perturbative 
low $\vec k_\sT^{\, 2}$ region. In Fig.~\ref{fig6}, all this is shown for 
$G_1^{\perp}$ in the same conditions and kinematics 
as in Fig.~\ref{fig5} (i.e., for $\xi=0.7$ and 
$\vec k_\sT \cdot \vec R_\sT =0$) and for $z_h=0.6,0.7,0.8,0.9$. It is evident 
that for increasing $z_h$, the fragmentation function gets concentrated 
at lower ${\vec k}_\sT^{\,2}$ and have an increasingly less important 
tail. This result can be generalized also to the other FF. In other
words, the more the hadron pair is leading, i.e. it takes most of the jet
energy, the more the FF are concentrated around the 
jet axis with smaller transverse momentum. 

The asymptotic behaviour of the FF for very large $\vec k_\sT^{\, 2}$ is 
dictated by the form factors at each vertex. It is instructive to look 
at the quark transverse momentum dependence for the contributions from each
diagram separately. From a simple power counting in propagators and vertices
(including form factors) in the diagrams, one can obtain the lower 
limit $\bar n$ in the $1/(\vec k_\sT^{\, 2})^n$ asymptotic 
decrease. Cancellations of leading power terms in the numerators can 
result in an even faster decrease of the actual results for 
the FF, i.e.~a higher power in the asymptotic 
$1/(\vec k_\sT^{\, 2})^n$ behavior. In Table~\ref{powertable} we 
compare $\bar n$ from each diagram with the actual asymptotic behavior of 
their contributions to the FF.

\begin{table}[hbt]
\begin{tabular}{|c|c|c|c|c|c|} 
$ $ & $\widetilde\Delta(3aS)$ & $\widetilde\Delta(3aA)$
& $\widetilde\Delta(4aA)$ & $\widetilde\Delta(4cS)$ 
& $\widetilde\Delta(4cA)$ \rule[-2mm]{0mm}{6mm}\\
\hline\hline
$\bar n$ & 3 & 2 & 4 & 4 & 3 \rule[-2mm]{0mm}{6mm}\\
\hline
$n(G_1^\perp)$     & - & - & 6 & 4.5 & 3.5 \rule[-2mm]{0mm}{6mm}\\
$n(H_1^\perp)$     & - & - & 5 & 4.5 & 3.5 \rule[-2mm]{0mm}{6mm}\\
$n(H_1^\newangle)$ & - & - & 5 & 4.5 & 3.5 \rule[-2mm]{0mm}{6mm}\\
\end{tabular} 
\caption{The lower limit $\bar n$ in the 
$1/(\vec k_\sT^{\, 2})^{\bar n}$ asymptotic decrease as obtained from a power 
counting. It is compared with the power actually found in the FF. The 
power law for the contributions from each diagram are shown separately, where 
$\widetilde\Delta((3/4)(a/c)(S/A))$ refers to the diagram (a/c) in 
Fig.~\ref{fig3}/\ref{fig4} involving the (scalar/axial) diquark.} 
\label{powertable}
\end{table}  

As already found in a different context~\cite{eikonal}, the lower limit
$\bar n$ is found to be smaller for the dominant ``pure'' channel related 
to the diagonal diagram~\ref{fig3}$a$, i.e.\ it shows the slowest asymptotic
fall-off in $\vec k_\sT^{\, 2}$. This agrees with the expectation from
dimensional counting rules which link the asymptotic behavior to the number of
outgoing partons in the lowest Fock state: 3 quarks in the Roper
compared to 5 (anti-) quarks in the proton-pion pair (the subsequent Roper
decay takes place spontaneously and does not affect the parton counting 
in this argument).

We have checked our results for stability under different choices for the 
quark mass parameter, and found a modest dependence. Even a drastic choice 
like setting the quark mass to zero results in less then 20\% 
changes of the FF functions (except for the extreme end-points of the 
kinematical regions where the results with zero and non-zero $m_q$ both 
vanish, and quoting of a relative size does not make sense). 

\section{Summary}
\label{sec:summary}

In this paper we have calculated the interference fragmentation functions 
(FF) that arise from the distribution of two hadrons produced in the same 
jet in the current fragmentation region of a hard process, e.g., in 
two-hadron inclusive lepton-nucleon scattering.

Naive ``T-odd'' FF generally arise because the existence of 
Final State Interactions (FSI) prevents constraints from time-reversal 
invariance to be applied. This class of FF is interesting because it allows
for a direct investigation of mechanisms for residual interactions inside 
jets and can contain the chiral-odd partners needed to isolate the
presently unknown quark transversity distribution $h_1$. 

The presence of FSI allows that in the fragmentation process there are at 
least two interfering competing channels. However, it has been
shown~\cite{noi} that this is not enough to generate ``T-odd'' FF. A
realistic microscopic description of the vertices involved in the
different channels is required which naturally selects two-hadron emission 
inside the same jet as the most convenient scenario.

To leading order, four FF arise, among which three are naive ``T-odd'' and
are related to the fragmenting quark being polarized longitudinally 
$(G_1^{\perp})$ or transversely $(H_1^{\perp}, H_1^{\newangle})$. The
latter two ones are, moreover, chiral odd and can be identified as the
partners to isolate $h_1$. In fact, asymmetry measurements in two-hadron 
inclusive DIS on a transversely polarized nucleon target induced by an 
unpolarized beam can be envisaged~\cite{noi}, where $H_1^{\newangle}$, 
$H_1^{\perp}$ 
enter the cross section in convolutions with $h_1$. In particular, 
$H_1^{\newangle}$ survives the integration over the transverse momentum of 
the fragmenting quark and can be deconvoluted from the transversity
distribution.

Calculations of such FF have been shown extending the diquark spectator 
model of Refs.~\cite{spectator,melni,jak97} and specializing it to 
the case of the 
hadron pair being a proton and a pion produced either directly or through 
the Roper resonance. Therefore, this is the first example of an explicit, 
complete and detailed model calculation of ``T-odd'' FF.

A well-known problem of this kind of approach is that it does not provide
a scale dependence. In principle, it is a ``low-scale model'', since it
involves simple valence quark degrees of freedom. In Ref.~\cite{jak97} a
comparison with parametrizations at ``low hadronic scale'' and available
experimental data has been attempted and a reasonable qualitative agreement
has been obtained for the known distribution functions and $D_1(z_h)$.
These findings give a good estimate for the values of the model
parameters, that here are also adopted to calculate the other unknown
fragmentation functions. Moreover, they give an indication of the level of
accuracy that can be reached disregarding sea-quarks, gluons and
evolution.

\acknowledgements{This work is part of the TMR program ERB FMRX-CT96-0008.
Interesting and fruitful discussion with P. Mulders are greatly 
acknowledged.}

\appendix

\section*{}
\label{sec:ApiA}

In this Appendix we give a justification for the Dirac structure, sign and
strength of the coupling adopted in Sec.~\ref{sec:three} for the (axial)
diquark-pion-(axial) diquark vertex $(A\pi A)$. In the following, we refer
to the same notations and conventions of Ref.~\cite{bd}.

The axial diquark is represented by a spin-1 field $\psi_{\mu}$ that
satisfies the equation of motion subject to the Lorentz condition, 
\begin{equation}
p^2 \psi_{\mu} = M_A^2 \psi_{\mu} \quad ; \quad 
p^{\mu} \psi_{\mu} = 0  \; , 
\label{eq:spin1}
\end{equation}
while the field tensor is defined as $F_{\mu \nu} = 
\partial_{\mu} \psi_{\nu} - \partial_{\nu} \psi_{\mu}$. The conserved 
vector current is given by 
\begin{equation}
V^{\mu} = i 
\left( (F^{\prime \mu \nu})^* \psi_{\nu} - F^{\mu \nu} \psi_{\nu}'^*\right) 
\; ,
\label{eq:vmu}
\end{equation}
where $F_{\mu \nu}' = \partial_{\mu} \psi_{\nu}' - \partial_{\nu}
\psi_{\mu}'$ and $\psi^{\left( ' \right)}$ refers to the incoming
(outgoing) spin-1 field in the vertex. Analogously, we define the diquark 
axial current
\begin{equation}
A^{\mu}_0 = f_{A\pi A} \  \epsilon^{\mu \alpha\beta\gamma} \left(
F'^*_{\beta\gamma} \psi_{\alpha} - F_{\beta \gamma} \psi'^*_{\alpha} \right)
\  = \  2 i f_{A\pi A} \  \epsilon^{\mu \alpha\beta\gamma} 
\left( \psi'^*_{\alpha}\  p_{\beta} \  \psi_{\gamma} 
- \psi'^*_{\gamma} \  p'_{\beta} \  \psi_{\alpha} \right) \; ,
\label{eq:amu0}
\end{equation}
where $p'$ refers to the momentum of the outgoing diquark. 
The $f_{A\pi A}$ is the weak axial-diquark coupling used in
Sec.~\ref{sec:three-b}, whose sign, however, is not yet determined, and the
definition of $F^{\mu \nu}$ in momentum space has been used. The current
$A^{\mu}_0$ does not fulfil the PCAC hypothesis. Analogously to the weak
neutron decay, we redefine the axial current introducing a pole term to
restore the PCAC hypothesis (or CAC, in case of vanishing pion mass
$m_{\pi}$), i.e.
\begin{equation}
A^{\mu} = A^{\mu}_0 + A^{\mu}_1 \equiv (g^{\mu \nu} -
\frac{q^{\mu} q^{\nu}}{q^2-m_{\pi}^2} ) A_{0 \nu} \; ,
\label{eq:amu}
\end{equation}
where $q^{\mu}$ is the momentum transferred from the vertex. The pole term
contains the diquark-pion-diquark $(A\pi A)$ vertex of interest. Its
explicit expression gives the Dirac structure of this vertex:
\begin{eqnarray}
A^{\mu}_1 &= &{}- \frac{q^{\mu} q^{\nu}}{q^2 - m_{\pi}^2} A_{0 \nu} \  = \  
- \frac{q^{\mu}\left( p^{\nu} - p'^{\nu} \right)}{q^2 -m_{\pi}^2} \  2i
f_{A\pi A} \  \epsilon_{\nu \alpha\beta\gamma} (\psi'^{* \alpha} p^{\beta}
\psi^{\gamma} - \psi'^{* \gamma} p'^{\beta} \psi^{\alpha} ) \nn \\
&= &4 \frac{q^{\mu}}{q^2 - m_{\pi}^2} \  \psi'^{* \alpha} \ i 
f_{A\pi A} \ \epsilon_{\nu\beta\alpha\gamma}\ p^{\nu} p'^{\beta} \ 
\psi^{\gamma} \  
\equiv \  4 \frac{q^{\mu}}{q^2 - m_{\pi}^2} \  \psi'^{* \alpha} \  
\Upsilon^{A\pi A}_{\alpha \gamma} \  \psi^{\gamma} \; .
\label{eq:amu1}
\end{eqnarray}

\begin{figure}[hbtp]
\begin{center}
\psfig{file=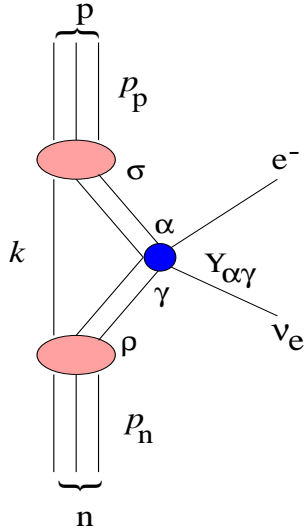, width=4cm, height=7cm}
\end{center}
\caption{\label{figB1} Weak decay of a neutron with momentum $p_n$ into
a proton with momentum $p_p$ via an axial diquark.}
\end{figure}

In order to determine whether $f_{A\pi A}$ is positive or negative, we need 
to explicitly calculate a diagram involving $A^{\mu}_0$, as in a 
hypothetical weak neutron decay $n \rightarrow p+e^- +\nu_e$ via an axial 
diquark depicted in Fig.~\ref{figB1}. Since the total neutron current has 
the $V-A \Leftrightarrow \gamma^{\mu} - \gamma^{\mu} \gamma_5 \equiv 
\gamma^{\mu} + \gamma_5 \gamma^{\mu}$ structure, 
our strategy will be to calculate $A^{\mu}_0$ from the diagram of 
Fig.~\ref{figB1} and then project out the $\gamma_5 \gamma^{\mu} $ part: 
the sign of the result will determine the sign of $f_{A\pi A}$ with respect 
to the vector current $V^{\mu}$. 

To calculate the diagram of Fig.~\ref{figB1} we use the rules
described in Sec.~\ref{sec:three} for the $qAp$ vertex, the $A\pi A$
vertex and the axial diquark propagator. The net result is a current
$G^{\mu}$ given by
\begin{eqnarray}
G^{\mu} &= &\int d^4 k \  \frac{1}{k^2 -m^2} \  \frac{1}{(p_p-k)^2-M_A^2}
\  \frac{1}{(p_n-k)^2-M_A^2} \nn \\
& &\quad \times \  2i f_{A\pi A} \ f_{qAp}^2 \ \gamma^{\sigma} \kslash 
\gamma^{\rho} \  \epsilon^{\mu \alpha\beta\gamma} \left[ (p_n-k)_{\beta} \  
g_{\alpha\sigma} g_{\rho\gamma} - (p_p-k)_{\beta} \  g_{\gamma\sigma}
g_{\rho\alpha} \right] \; .
\label{eq:gmu}
\end{eqnarray}
Moreover, the Dirac-structure decomposition of a matrix $M$ contains the
$\gamma_5 \gamma^{\mu} $ projection as 
\begin{equation}
M = \ldots - \frac{1}{4} {\rm Tr} [ \gamma_5 \gamma^{\mu} M] \  \gamma_5 
\gamma^{\mu} \ldots \; .
\label{eq:decomp}
\end{equation}
Therefore, the $\gamma_5 \gamma^{\nu}$ projection of $G^{\mu}$ will be 
\begin{equation}
\begin{array}{l} - \frac{1}{4} {\rm Tr} [ \gamma_5 \gamma^{\nu} G^{\mu} ] = 
\displaystyle{\int d^4 k \  \frac{1}{k^2 -m^2} \  
\frac{1}{(p_p-k)^2-M_A^2} \  \frac{1}{(p_n-k)^2-M_A^2}} \\[5mm]
\quad \begin{array}{l} \times \  \left\{ - 4 f_{A\pi A} \  f_{qAp}^2 \  
\left[ k^{\mu} \  (p_n+p_p-2k)^{\nu} - k\cdot (p_n+p_p-2k) \  g^{\mu \nu} 
\right] \right\} \end{array} \end{array} \; .
\label{eq:projgmu}
\end{equation}
To determine the sign of the r.h.s.\ of Eq.~(\ref{eq:projgmu}), let us
consider the component $\mu=\nu=0$ only. Noting that strong off-shellness
of the particles is suppressed by the form factors, propagators in 
Eq.~(\ref{eq:projgmu}) are dominated by their positive energy poles. Writing
the quark energy as $k^0 = m+\epsilon$ with $0<\epsilon \ll m$, and taking
$m_p = m_n \sim 3m$ and $M_A \sim 2m$, we finally get 
\begin{equation}
- \frac{1}{4} {\rm Tr} \left[ \gamma_5 \gamma^0 G^0 \right] \sim - 8 \  
f_{A\pi A} \  f_{qAp}^2 \  m \epsilon \; .
\label{eq:projg0}
\end{equation}
Keeping in mind the $V-A\Leftrightarrow \gamma^{\mu} + \gamma_5 
\gamma^{\mu}$ structure of the current, we deduce that $f_{A\pi A}$ is a
negative quantity and is parametrized as 
\begin{equation}
f_{A\pi A}(P_A^2) = - N_{\pi A} \; 
   \frac{P_A^2 -M_A^2}{|P_A^2-\Lambda_{\pi}^2|^{\alpha}} \; .
\label{eq:ApiA}
\end{equation}

The last open problem is to determine the strength of this coupling, in
other words the size of $N_{\pi A}$. We can assume that the pion emission,
as it would occur in a pion exchange between, e.g., two protons, is of the 
same order for all constituents of the protons, irrespective of modelling the 
proton as an ensemble of three valence quarks or of a quark 
and a diquark. Therefore, we can compare the amplitudes corresponding to the
quark-pion-quark vertex, ${\cal M}_{q\pi q}$, and to the diquark-pion-diquark
vertex, ${\cal M}_{A\pi A}$, assuming that they represent part of the
diagrams describing the pion exchange between the constituents (quark,
diquark) of a proton and another particle. Using again the rules defined in
Sec.~\ref{sec:three} we can write
\begin{eqnarray}
\vert {\cal M}_{q\pi q} \vert^2 &= &{\rm Tr} \left[ f_{q\pi q} \gamma_5
\frac{i}{\kslash '-m} f_{q\pi q} \gamma_5 \frac{i}{\kslash -m} \right] \nn
\\
& &= 4 \frac{f_{q\pi q}^2}{(k'^2-m^2)(k^2-m^2)} (k\cdot k' -m^2) \nn
\\[.3cm]
\vert {\cal M}_{A\pi A} \vert^2 &= &{\rm Tr} \left[ -if_{A\pi A}\  
\epsilon_{\mu \nu \alpha \beta} \  k^{\alpha} k'^{\beta} \  \frac{i}{k'^2-M_A^2}
(-g^{\nu \rho}) (-i f_{A\pi A}) \  \epsilon_{\rho\sigma\alpha'\beta'}\  
k'^{\alpha'} k^{\beta'} \  \frac{i}{k^2-M_A^2} (-g^{\sigma\mu}) \right] \nn \\
& &=2 \frac{f_{A\pi A}^2}{(k'^2-M_A^2)(k^2-M_A^2)} \left[ (k\cdot k')^2
-M_A^4 \right] \; ,
\label{eq:ampl}
\end{eqnarray}
where $k^{(')}$ is the incoming (outgoing) momentum in the vertex for both
quark and diquark and the couplings $f_{q\pi q}, f_{A\pi A}$ are assumed
pointlike for simplicity. If the virtuality of the pion is small, in the
c.m.\ of, e.g., the initial quark with $k=(m,\vec 0)$ we can assume $k\cdot
k' \simeq m \epsilon$, with $\epsilon \sim m$ some energy such that
$\epsilon -m \ll m$. Similar arguments hold also for the diquark and are
independent of the angular distribution of the pion. Inserting these
relations in Eq. (\ref{eq:ampl}), we get 
\begin{equation}
\vert {\cal M}_{A\pi A} \vert^2 \sim \vert {\cal M}_{q\pi q} \vert^2 \;
\Longrightarrow \; f_{A\pi A} \sim \frac{f_{q\pi q}}{M_A} \; .
\label{eq:strength}
\end{equation}
If the couplings are not pointlike and have the structure shown in
Sec.~\ref{sec:three-b}, we need to search for a scale $k_0^2$ at which they
are comparable functions of $k^2$, i.e.
\begin{equation}
|f_{A\pi A}(k_0^2)| = N_{\pi A} \frac{k_0^2 -M_A^2}{\vert
k_0^2-\Lambda_{\pi}^2\vert^2} \sim \frac{|f_{q\pi q}(k_0^2)|}{M_A} =
\frac{N_{q\pi}}{M_A} \frac{k_0^2-m^2}{\vert
k_0^2-\Lambda_{\pi}^2\vert^{3/2}} \; ,
\label{eq:comp}
\end{equation}
with $\Lambda_{\pi} = 0.4$ GeV and $N_{q\pi} = 2.564$ GeV~\cite{jak97}. This
is true for $N_{A\pi A} = 6$ GeV and $1<k_0^2<2$ GeV$^2$, which is
reasonably far from the pole $\Lambda_{\pi}$ and from the asymptotic scale
where the two form factors have very different power laws.

\end{document}